\documentclass[aps,prd,reprint,superscriptaddress,nofootinbib]{revtex4-2}
\usepackage[english]{babel}
\usepackage[utf8]{inputenc}
\usepackage[colorinlistoftodos, color=green!40, prependcaption]{todonotes}
\usepackage{orcidlink}
\usepackage{amsthm}
\usepackage{amsmath}
\usepackage{mathtools}
\usepackage{physics}
\usepackage{slashed}
\usepackage{xcolor}
\usepackage{graphicx}
\usepackage{adjustbox}
\usepackage{placeins}
\usepackage[T1]{fontenc}
\usepackage{comment}
\usepackage{lipsum}
\usepackage{csquotes}
\usepackage{hyperref}
\usepackage{lipsum}
\usepackage[normalem]{ulem}
\usepackage{subfigure}
\usepackage{caption}
\usepackage{subcaption}
\usepackage{amssymb}
\begin{document}
\title{Probing the Sterile Neutrino Dipole Portal with SN1987A \\ and Low-Energy Supernovae}
\author{Garv Chauhan\,\orcidlink{0000-0002-8129-8034}}
    \email{gchauhan@vt.edu}
    \affiliation{Center for Neutrino Physics, Department of Physics, Virginia Tech, Blacksburg, VA 24061, USA}
\author{Shunsaku Horiuchi\,\orcidlink{0000-0001-6142-6556}}
    \affiliation{Center for Neutrino Physics, Department of Physics, Virginia Tech, Blacksburg, VA 24061, USA}
    \affiliation{Kavli IPMU (WPI), UTIAS, The University of Tokyo, Kashiwa, Chiba 277-8583, Japan}
\author{Patrick Huber}
    \affiliation{Center for Neutrino Physics, Department of Physics, Virginia Tech, Blacksburg, VA 24061, USA}
\author{Ian M. Shoemaker\,\orcidlink{
0000-0001-5434-3744}}
    \affiliation{Center for Neutrino Physics, Department of Physics, Virginia Tech, Blacksburg, VA 24061, USA}

\begin{abstract}
BSM electromagnetic properties of neutrinos may lead to copious production of sterile neutrinos in the hot and dense core of a core-collapse supernova. In this work, we focus on the active-sterile transition magnetic moment portal for heavy sterile neutrinos. Firstly, we revisit the SN1987A cooling bounds for dipole portal using the integrated luminosity method, which yields more reliable results (especially in the trapping regime) compared to the previously explored via emissivity loss, aka the Raffelt criterion. Secondly, we obtain strong bounds on the dipole coupling strength reaching as low as $10^{-11} \text{ GeV}^{-1}$ from energy deposition, i.e., constrained from the observation of explosion energies of underluminous Type IIP supernovae. In addition, we find that sterile neutrino production from Primakoff upscattering off of proton dominates over scattering off of electron for low sterile neutrino masses. 
\end{abstract}

\maketitle

\section{Introduction} 
Neutrino flavor oscillations imply that neutrino masses are nonzero, a fact not accounted for in the Standard Model (SM). However, the observation of nonzero neutrino masses can be explained if the SM is augmented with at least two right-handed sterile neutrinos (for the two mass-splittings). In the absence of firm experimental guidance, we do not know how heavy, how many, or how interacting these sterile neutrinos are. As a result, a broad multi-scale experimental and observational program is underway~\cite{Abdullahi:2022jlv}.

The most studied phenomenological set-up for sterile neutrinos is to assume that their mass-mixing parameters are the keys to their production as well as detection. This is not however the only possibility. For example, there are well-motivated scenarios in which a relatively large transition dipole moment between active and sterile neutrinos dominates their behavior (e.g.~\cite{Dolgov:2002wy,Coloma:2017ppo,Plestid:2020vqf,Brdar:2020quo,Brdar:2023tmi}). A large phenomenological program has ensued to constrain active-sterile dipole moments by making use of an array of terrestrial, astrophysical, and cosmological data~\cite{Gninenko:2009ks,Gninenko:2010pr,McKeen:2010rx,Masip:2011qb,Masip:2012ke,Coloma:2017ppo,Magill:2018jla,Brdar:2020quo,Shoemaker:2018vii,Arguelles:2018mtc,Fischer:2019fbw,Coloma:2019htx,Schwetz:2020xra,Arina:2020mxo,Shoemaker:2020kji,Abdullahi:2020nyr,Shakeri:2020wvk,Atkinson:2021rnp,Cho:2021yxk,Dasgupta:2021fpn,Arguelles:2021dqn,Ismail:2021dyp,Miranda:2021kre,Bolton:2021pey,Jodlowski:2020vhr,Vergani:2021tgc}. Lastly, we note that the possibility of neutrinos having nonzero magnetic moments has a long history, going back to Pauli's letter in 1930 in which the neutrino was proposed as a new particle~\cite{Pauli:1930pc}. 

To date, some of the most sensitive probes of active-sterile dipole moments have involved supernovae (SNe)~\cite{Magill:2018jla,Brdar:2023tmi}. If their production is too frequent, they can lead to excessive cooling of SN1987A~\cite{Magill:2018jla}, or produce an overabundance of detectable neutrinos or photons~\cite{Brdar:2023tmi}. However recently, low-energy supernovae have emerged as powerful probes of new physics~\cite{Caputo:2022mah,Chauhan:2023sci}. In this paper, we will derive new constraints on active-sterile dipole moments from deposition of excess energy in low-energy supernovae, which is constrained from the observations of  SN Type IIP light curves. We also re-visit the SN1987A bounds in light of additional production modes, finding important differences with existing literature. 

This paper is organized as follows. In Section~\ref{sec:dipoleatSN} we describe the various production modes of sterile neutrinos via the dipole interaction, and compute their luminosity as a function of their mass and dipole coupling. In Section~\ref{sec:bounds} we discuss the observational constraints from SNe that allow us to impose constraints on active-sterile dipole moments. Finally in Sec.~\ref{sec:results} we display our main results and discuss them in the context of the existing constraints on the dipole portal.

\section{Dipole Portal at Supernovae} 
\label{sec:dipoleatSN}

After electroweak symmetry breaking, the effective Lagrangian for the dipole portal involving active-sterile transition magnetic moment can be written as
\begin{equation}
\mathcal{L} \, \supset \, i\bar{N} \slashed{\partial}N + \sum_\alpha d_{\alpha} \bar{N} \sigma_{\mu\nu} \nu^\alpha_L F^{\mu\nu} - \frac{M_N}{2}\bar{N}^{c}N + h.c.
\label{eq:lagr}
\end{equation}
where $N$ is a sterile neutrino, $\nu_L$ is a SM left-handed neutrino field, $F^{\mu\nu}$ is the electromagnetic field strength tensor, and $d_\alpha$ is the active-sterile transition magnetic moment. We assume the coupling strength to be flavor universal, i.e., $d_\alpha=d$ . {For specific UV scenarios explaining the origin of this coupling, see, e.g., Refs.~\cite{Aparici:2009fh,Coloma:2017ppo,Babu:2020ivd,Brdar:2020quo,Ismail:2021dyp}.}

\subsection{Production} 

\begin{figure*}[t]
\centering
  \includegraphics[width=1.99 \columnwidth]{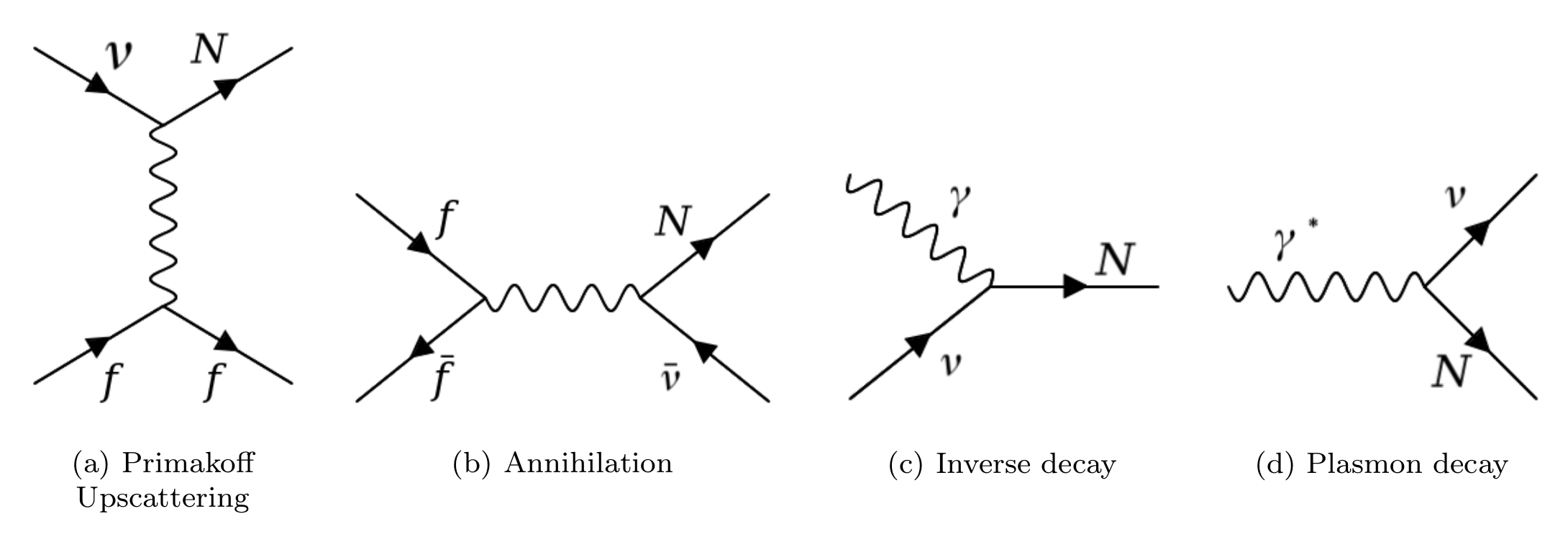}
\caption{Sterile neutrino production modes in SN through dipole portal via (a) Primakoff upscattering off a charged fermion via photon exchange, (b) charged fermion annihilation to a sterile and active neutrino, (c) photon + neutrino inverse decay, and (d) plasmon decay. }
\label{fig:prodfigs}
\end{figure*}

For a given active-sterile neutrino transition magnetic moment, heavy sterile neutrinos can be  produced in a SN core through neutrino scattering off of electrons $e^{\pm}$, muons $\mu^{\pm}$ and protons $p$, through pair-annihilation of $e^{\pm}$ or $\mu^{\pm}$, inverse decay and through plasmon decay (see Fig.~\ref{fig:prodfigs}). Despite the high number density, neutrons do not play any role in sterile neutrino production at the tree level. The relevant production modes are listed below~\cite{Magill:2018jla}:
\begin{align}
    \nu + p  &\rightarrow N + p, &\qquad\text{(upscattering)} \label{eq:mode1}\\
    \nu + e^\pm &\rightarrow N + e^\pm,  &\qquad\text{(upscattering)}  \\
    \nu + \mu^\pm &\rightarrow N + \mu^\pm, &\qquad\text{(upscattering)}  \\
    e^+ + e^- &\rightarrow \bar{\nu} + N, &\qquad\text{(annihilation)}  \\
    \mu^+ + \mu^- &\rightarrow \bar{\nu} + N,  &\qquad\text{(annihilation)}  \\
    \nu + \gamma &\rightarrow N &\qquad \text{(inverse decay)}  \\
    \gamma^* &\rightarrow N + \Bar{\nu} &\qquad\text{(plasmon decay)} \label{eq:mode7} 
\end{align}
The matrix elements for these processes have been calculated and provided in the appendix. In this work, we significantly improve on the production rate calculation in the literature, by including the effect of muon population, plasmon decay channel and the gravitational effects of the high-density proto-neutron star core. We also discuss and highlight a major result of our work: the dominance of neutrino upscattering off of proton over upscattering through electron for low $M_N$. 

Primakoff upscattering occurs through a $t$-channel exchange of a photon with the SN medium composed of protons, electrons and muons.
However, as can be seen in the matrix element for this process in Eq.~(\ref{eq:Mprim}) prefers strong forward scattering. In vacuum, this diagram is regulated by restricting the angular range to forward scattering angles determined by the minimum momentum transfer required for sterile neutrino production in the final state \cite{Raffelt:1996wa,Plestid:2020vqf,Brdar:2020quo}. However, in presence of a medium, the photon develops a non-trivial dispersion relation acquiring an effective plasmon mass, which can help regulate the total cross-section. The effective mass of the transverse photon modes generally is of $\mathcal{O}{(\omega_\text{P})}$, i.e., the plasma frequency. Including the contributions from electrons and protons in the SN medium respectively, $\omega_\text{P}$ is given by
\begin{equation}
    \omega_\text{P}^2 = \frac{4\alpha}{3\pi}\left( \mu_e^2 + \frac{\pi^2 T^2}{3}\right) + \frac{4\pi\alpha\,n_p}{m_p},
    \label{eq:thermalMass}
\end{equation}
where $\alpha$ is the fine-structure constant, $\mu_e$ is the electron chemical potential, $T$ is the temperature of the SN core, and $n_p$ and $m_p$ are the number density and mass of the proton, respectively. Due to the high $\mu_e$ and high $m_p$ ($\gg T$), $\omega_\text{P}$ is usually dominated by the relativistic electron plasma frequency (i.e., the 1st term). For typical $\mu_e \sim 250$ MeV, $\omega_\text{P}$ usually is of $\mathcal{O}(10$ MeV). 

In addition, there is another screening length $k_\text{S}$ determined by the \textit{Debye-H\"uckel} scale $k_\text{D}$ for non-degenerate non-relativistic medium and by the \textit{Thomas-Fermi} scale $k_{\text{TF}}$ for degenerate medium. It arises from the movement of charged species in the medium, leading to charge screening of the target. The net screening scale including contributions from the proton and electrons respectively, is given by 
\begin{equation}
    k_\text{S}^2 = \frac{4\pi\alpha}{T}n_p + \frac{4\alpha}{\pi}\left( \mu_e^2 + \frac{\pi^2 T^2}{3}\right),
    \label{eq:debyeMass}
\end{equation}
where $n$ denotes the number density of protons. Note that $k_\text{S}$ doesn't suffer any suppression from the proton mass as compared to $\omega_\text{P}$. Since highly degenerate and relativistic electrons in the SN core forms a stiff background, the dominant contribution to $k_\text{S}$ comes from protons and other heavy ions. This can also be seen from Eq.~\eqref{eq:debyeMass}, since $n_p \simeq n_e$ (charge neutrality) and $n_e=\mu_e^3/3\pi^2$ (degenerate fermi gas), the electron contribution in the second term is suppressed by a factor of $T/\mu_e$.

From Eqs.~\eqref{eq:thermalMass} and~\eqref{eq:debyeMass}, we can clearly see that $\omega_\text{P} < k_\text{S}$, i.e., charge screening tends to be the dominant scale. Hence, ignoring $\omega_P$ and considering photons to be massless is a good approximation for processes involving scattering off of charged targets and $k_\text{S}$ can help regulate the $t$-channel singularity. To include this screening effect for the Primakoff upscattering process, we make the following change to the matrix element,
\begin{equation}
     |\mathcal{M}|^2 \rightarrow
    |\mathcal{M}|^2 \frac{q^4}{(q^2-k_\text{S}^2)^2},
    \label{eq:subM}
\end{equation}
where $q^2$ is the 4-momentum carried by the photon propagator. Although the more apt substitution is $ |\mathcal{M}|^2 \rightarrow |\mathcal{M}|^2 \frac{\textbf{q}^2}{(\textbf{q}^2+k_\text{S}^2)}$, where $\textbf{q}$ is the 3-momentum of the photon in the rest frame of the medium. Although in absence of complete thermal rates available for all production modes, we stick to the easier substitution in Eq.~(\ref{eq:subM}). Previously in the literature \cite{Magill:2018jla}, a lower cutoff on $q^2$ was used, which is essentially equivalent to including a Debye screening effect in the matrix element, as shown in Eq.~(\ref{eq:subM}). 

For any scattering involving the proton, the Dirac form factor $F_1(q^2)$ needs to be taken into account. We provide the relevant nuclear charge form factor in Appendix~\ref{sec:appendixPrim}, although for most $q^2$ of interest in our case, $F_1(q^2)\simeq 1$. Note that in this work, we neglect the effect of nucleon magnetic moments and will be included in a future study including the thermal effects for Primakoff upscattering. 

The production through annihilation $f\Bar{f}\rightarrow N\Bar{\nu}$, where $f=e,\mu$, is shown in Fig.~\ref{fig:prodfigs}(b). 
Due to the $s$-channel exchange of a photon, this process does not suffer from the ``forward'' scattering issue encountered for Primakoff upscattering. Since there is also no scattering off charged species involved, the effect of the screening scale $k_\text{S}$ is absent. The  $|\mathcal{M}|^2$ for this process can be obtained by applying crossing symmetry rules to the (vacuum) matrix element for the Primakoff upscattering given in Eq.~(\ref{eq:Mprim}).

Since the photons and neutrinos are thermalized in the SN core, the $N$ production can also proceed through inverse decays $\gamma\nu\rightarrow N$ (See Fig.~\ref{fig:prodfigs}(c)). The matrix element for this process is given in Eq.~\eqref{eq:matrixInv}. Usually, $M_N$ up to $\sim 6 T$ is accessible but for $\nu_e$ with high chemical potential $\mu\gg 3T$, heavier $N$s can also be produced without significant Boltzmann suppression.  

As discussed earlier, due to interactions with a high temperature and density medium, photons develop a thermal mass. Thus, the decays of photons also become kinematically allowed in a SN core, as shown in Fig.~\ref{fig:prodfigs}(d). In our case, this mode is important only for sterile masses $M_N\lesssim \omega_P$. The decay rate is given in Eq.~\eqref{eq:GammaPlasmon} and detailed production rates are discussed later.   

\subsection{Boltzmann Equations} 
The simplified kinetic equation for sterile neutrino production is,
\begin{equation}
    \frac{\partial f_N}{\partial t} = \mathcal{C}_{coll}(f_N),
\end{equation}
where $f_N$ is the sterile neutrino phase-space density distribution and $\mathcal{C}_{coll}$ is the sum of all possible collisional interactions. In our case, $\mathcal{C}_{coll}$ includes $2\rightarrow2$, $2\rightarrow1$ and $1\rightarrow2$ processes. The collisional term for $2\rightarrow 2$ particle interactions can be written \cite{Mastrototaro:2019vug,Hannestad:1995rs,Mastrototaro:2021wzl,Hahn-Woernle:2009jyb, Tamborra:2017ubu},
\begin{align}
    \mathcal{C}_{coll}(f_N) = &\frac{1}{2 E_N} \int d^3 \Tilde{p_2} d^3 \Tilde{p_3} d^3 \Tilde{p_4}\, \Lambda(f_N,f_2,f_3,f_4) \times \nonumber \\ &  |M|^2_{12\rightarrow 34}\, \delta^4(p_N+p_2-p_3-p_4) (2 \pi)^4,
    \label{eq:collint}
\end{align}
where $d^3 \Tilde{p_i}= d^3 {p_i}/((2 \pi^3)\,2 E_i)$, $\Lambda(f_N,f_2,f_3,f_4)= (1-f_N)(1-f_2)f_3 f_4- f_N f_2(1-f_3)(1-f_4)$ is the phase-space factor including the Pauli blocking of final states, $|M|^2$ is the interaction matrix element element squared including the symmetry factor, and $E_i$ and $p_i$ are energy and momentum of the $i$-th particle. The collisional integrals for  $2\rightarrow1$ and $1\rightarrow2$ can be obtained similarly (See Appendices~\ref{sec:appendixInv} and~\ref{sec:appendixPlasm}). 

For the $N$ production rate, we assume the dipole strengths are weak enough to not affect the standard SN processes. We also set the initial distribution $f_N=0$, since for such range of $|d|$, the sterile neutrino produced will not be trapped and thermalized in the SN. After solving for $f_N$, we can calculate the differential luminosity as~\cite{Tamborra:2017ubu,Mastrototaro:2019vug},
\begin{equation}
    \frac{dL_N}{dE_N} = \frac{2 E_N}{\pi} \int dr\,r^2 \frac{df_N}{dt} E_N\,p_N.
    \label{eq:diff} 
    \end{equation}

While the distribution functions for the leptons ($l$) have the usual Fermi-Dirac form determined by $p_l,m_l,\mu_l$ and $T$, the case for nucleons is quite different due to strong interactions under high densities leading to the breakdown of non-interacting picture. The mean-field potentials arising from nucleon self-energies play an important role. In our case, they modify the dispersion relation for nucleons and significantly effect their Pauli-blocking factors. The dispersion relation for nucleons, considering them as a non-relativistic quasi-particle gases moving under a mean-field potential $U$, is given \cite{Martinez-Pinedo:2012eaj,Mirizzi:2015eza},
\begin{equation}
    E(\textbf{p}) = \frac{\textbf{p}^2}{2m^*} + m + U,
\end{equation}
where $m$ and $m^*$ are the rest mass and Landau effective masses of the nucleon, respectively. $m^*$ and $U$ are both functions of temperature, density and the neutron-to-proton ratio. Given the nucleon chemical potential (with rest mass included), we can now define the nucleon distribution function as, 
\begin{equation}
    f_{\text{nucleon}}(p)=\frac{1}{\text{exp}\left[\frac{\sqrt{p^2+{m^*}^2}-\mu^*}{T} \right]+1},
\end{equation}
where we define the effective nucleon chemical potential $\mu^*=\mu-U$.

\begin{figure}[htb!]
    \centering
    \includegraphics[width=\columnwidth]{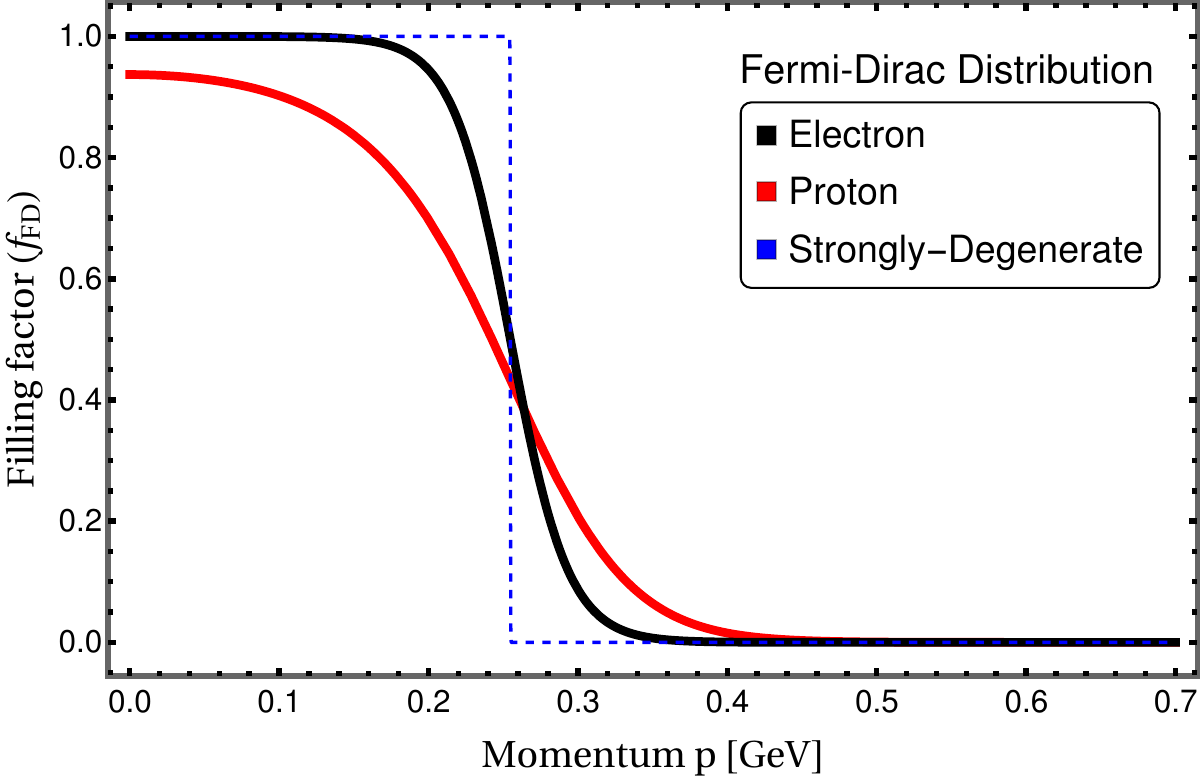}
    \includegraphics[width=\columnwidth]{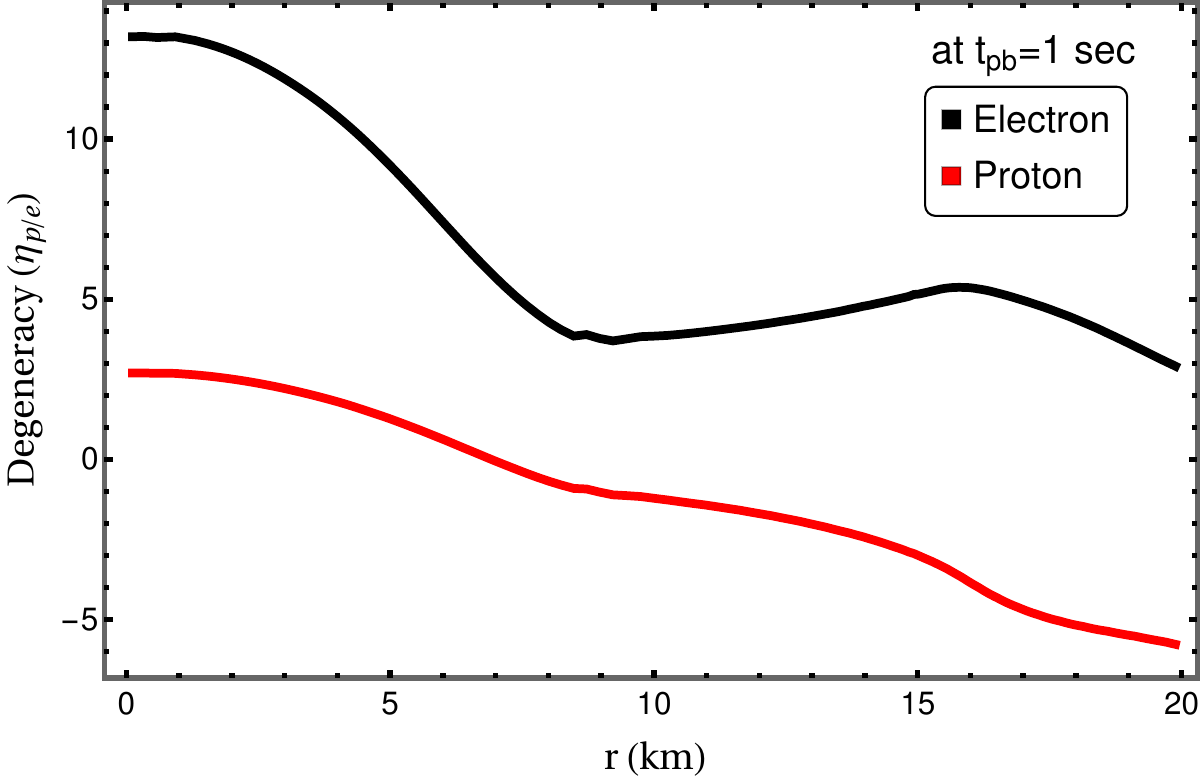}
    \caption{(Upper panel) Fermi-Dirac distribution for three cases: proton, electron and strongly degenerate gas at $r=2$ km ($t_{pb}=1$ sec). (Lower panel) Degeneracy parameter $\eta_{deg}$ for $e$ and $p$ for our SN profile at $t_{pb}=1$ sec.}
    \label{fig:p3}
\end{figure}

We can now define a useful concept for later discussions to quantify the degeneracy of Fermi gases. A Fermi gas is strongly degenerate when the chemical potential is greater than the average thermal energy. Therefore, the degeneracy parameter $\eta_{deg}$ is defined as,
\begin{equation}
    \eta_{deg} = \frac{\mu-m}{T}.
\end{equation}
Note for nucleons, we replace $\mu\rightarrow \mu^*$ and $m\rightarrow m^*$. Thus, $\eta_{deg}\gg1$ is strongly degenerate while $\eta_{deg}<0$ is non-degenerate. For example, $\eta_{deg}$ for the SN profile used in this work at post-bounce time $t_{pb}=1$ sec is shown in Fig.~\ref{fig:p3} (lower panel). While the electrons are strongly degenerate at all radii inside the SN core, the protons are only slightly degenerate in the center and  turn non-degenerate at $r>6$ km. The upper panel in Fig.~\ref{fig:p3} shows the filling factor for the momentum states for electrons and protons at $r=2$ km ($t_{pb}=1$ sec). We also include the case of strongly degenerate gas for comparison, assuming $\mu=\mu_e$ and $T\simeq0$. Degeneracy has strong effects on the production rate. For example, the presence of highly-degenerate species like electrons in the final state can suppress the production rate compared to the non-degenerate protons. 

\begin{figure}[ht!]
    \centering
    \begin{subfigure}
        \centering
        \includegraphics[width=\columnwidth]{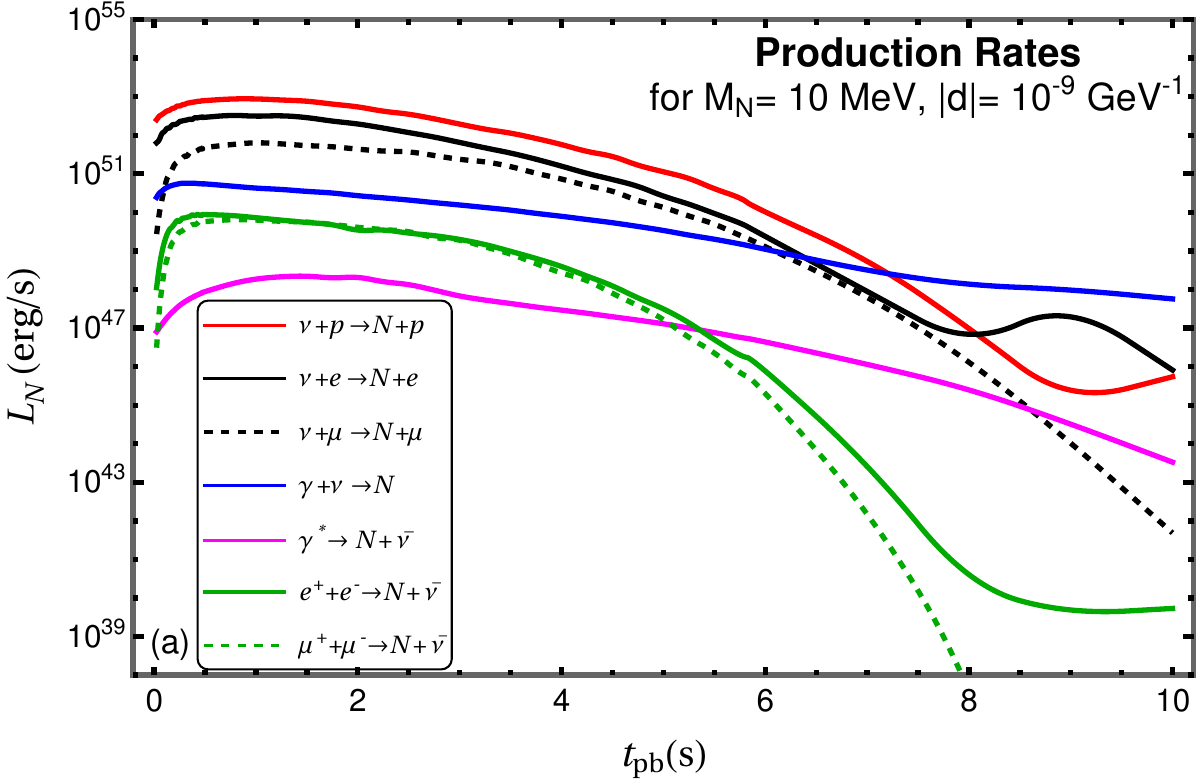}
    \end{subfigure}%
    \begin{subfigure}
        \centering
        \includegraphics[width=\columnwidth]{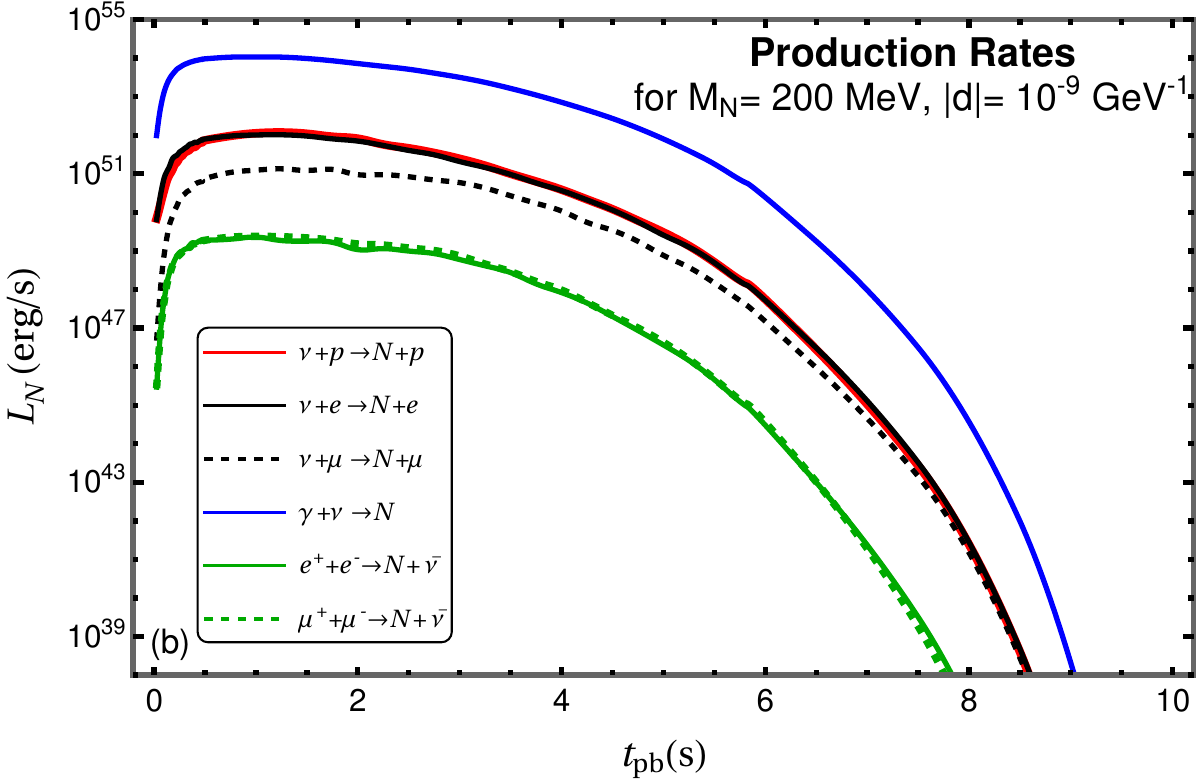}
    \end{subfigure}%
    \begin{subfigure}
        \centering
        \includegraphics[width=\columnwidth]{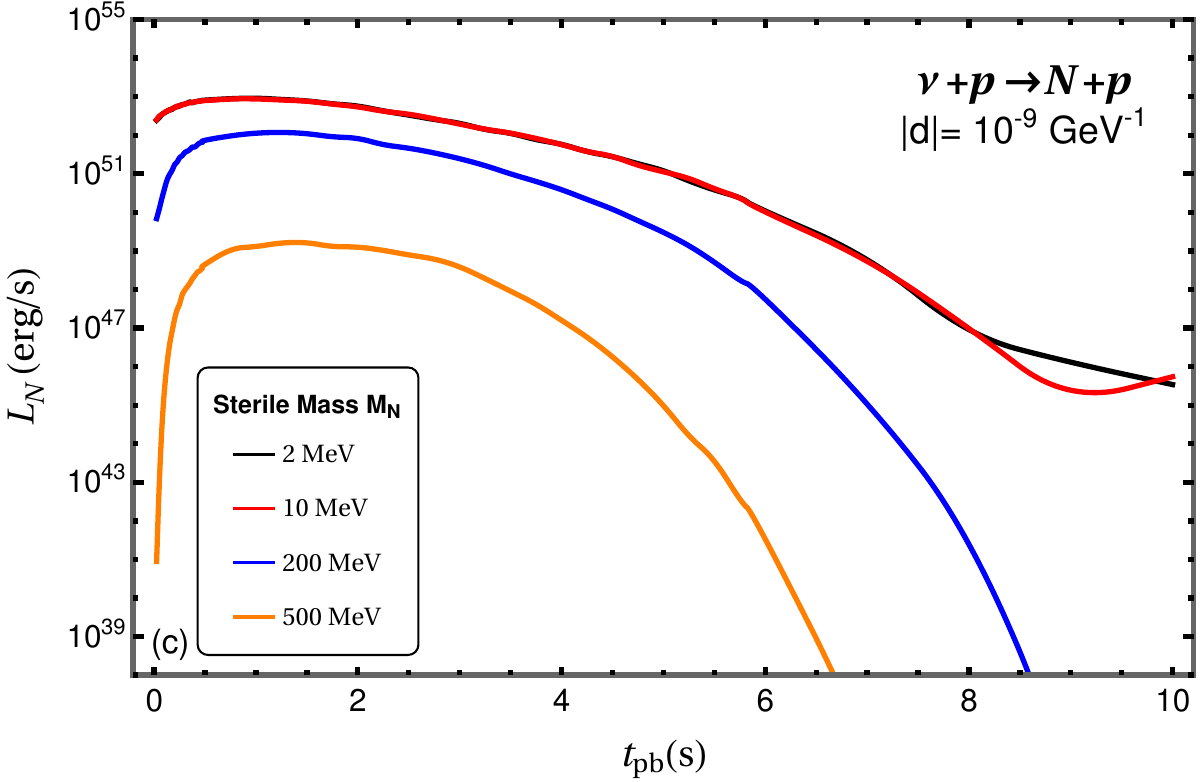}
    \end{subfigure}
    \caption{Sterile neutrino luminosity as a function of time for $|d|=10^{-9} \text{ GeV}^{-1}$ for (a) all production modes at $M_N=10$ MeV, (b) all production modes at $M_N=200$ MeV, and (c) proton Primakoff upscattering mode for different $M_N$.}
    \label{fig:prod}
\end{figure}

For the SN profile used in this work (details in the next section), Fig~\ref{fig:prod} shows the different contributions to sterile neutrino luminosity $L_N$ as functions of time. In Fig~\ref{fig:prod}(a), $L_N$ is shown for all production modes listed in Eqs.~\eqref{eq:mode1}-\eqref{eq:mode7} for $|d|=10^{-9} \text{ GeV}^{-1}$ at $M_N=10$ MeV. The proton Primakoff $\nu p \rightarrow N p$ is the dominant process for $M_N=10$ MeV, with the rate of electron Primakoff $\nu e \rightarrow N e$ following closely. Despite the same number densities as required by charge neutrality, the difference between the rates can arise from the high degeneracy of electrons, which lead to suppression of the production rate as compared to the proton case. The muon Primakoff is further suppressed due to the lower number density of muons, i.e., $n_\mu <n_e,n_p$. The rate from plasmon and inverse decay processes, although sub-dominant to Primakoff scattering, does not fall off as strongly as the kinematic limit is enhanced from the high-chemical potential of $\nu$'s and due to the absence of Pauli blocking. In fact, even after chemical potentials drop between 8-10 sec, the average thermal energy of $\nu$ in $\nu\gamma \rightarrow N$ is sufficient for $N$ production for low $M_N$. The production rate from annihilation channels $f\Bar{f} \rightarrow N \nu$ are mainly determined by the chemical potentials $\mu_e$ and $\mu_\mu$. It can be seen from the SN profile that $\mu_e\sim \mu_\mu$ for most $t_{pb}$, thereby leading to the same production rate at most times. The overall magnitude of the annihilation rate is suppressed compared to the Primakoff process, due to the suppressed number density of anti-fermions.

Similarly, in Fig~\ref{fig:prod}(b) $L_N$ is shown for all relevant production modes for $M_N=200$ MeV for $|d|=10^{-9} \text{ GeV}^{-1}$. For heavier steriles, essentially all production modes will suffer severe Boltzmann suppression, especially at later times since temperatures and chemical potentials have dropped significantly by then. The rate for proton and electron Primakoff upscattering are quite similar (notice the log-scale for $L_N$) since the heavy sterile production cannot just proceed through scattering off the Fermi surfaces only \footnote{For light $N$, the initial $f$ state has $E_{f}\sim\mu_e$ and can be placed back on the Fermi surface in the final state $\rightarrow f_{\text{nucleon}}=1/2$, leading to no exponential suppression from degeneracy.} and suppression from high degeneracy leads to exponential suppression. This also explains why the inverse decay dominates in this case. Since typical $\omega_P \ll 200$ MeV, the plasmon mode is absent in this case. Similar to the low $M_N$ case, annihilation channels have nearly the same rates due to similar chemical potentials $\mu_e$ and $\mu_\mu$. 

In Fig~\ref{fig:prod}(c), $L_N$ is shown for only the proton Primakoff upscattering process for different values of $M_N$ at fixed $|d|=10^{-9} \text{ GeV}^{-1}$.

\section{Supernovae Bounds}
\label{sec:bounds}
We discuss two different methods to obtain bounds on the dipole portal physics using SNe: 
(i) Raffelt criterion, and (ii) Integrated Luminosity (IL) criterion. While the former is a locally derived constraint on the energy lost by production of new particles, the latter is a global one. 

The Raffelt criterion is applied at a characteristic radius and requires the local emissivity of the sterile neutrinos at $r_0$ to not exceed more than 10$\%$ of the total neutrino emissivity ~\cite{Raffelt:1996wa,Dreiner:2003wh,Dreiner:2013mua,Magill:2018jla}, i.e., 
\begin{equation}
    \frac{\text{d}\varepsilon_N}{\text{d}t}(r_0) \leq \frac{1}{10} \frac{\text{d}\varepsilon_\nu}{\text{d}t} \simeq \frac{\rho(r_0)}{{\rm g}/{\rm cm}^3}\times 10^{19}\, {\rm erg}\,{\rm cm}^{-3}\,{\rm s}^{-1}.
\end{equation}
For the integrated luminosity criterion, the energy-loss rate per unit mass can be converted to a total luminosity loss by taking the mass of the SN core and the duration of the SN event into account. Observations of energy-loss rate from SN1987A, assuming $M_{core}\sim 1 M_\odot$, leads to the following upper bound 
\begin{equation}
    E_{N,{\rm cool}} < 10^{52}\,  {\rm erg}.
\end{equation}
Another class of constraint from SNe stems from the identification of a sub-class of SNe with low explosion energies, termed underluminous Type IIP SNe. These have been recently used to constrain the parameter space of axions~\cite{Caputo:2021rux} and sterile neutrinos~\cite{Chauhan:2023sci,Carenza:2023old}. The explosion energy released in SNIIP explosions can be inferred from the spectrum and light curves. Using fitting formulae, simulations, and statistical inference, the lowest SNIIP explosion energies inferred is some $7.4 \times 10^{49}$ \cite{Pejcha:2015pca,Muller:2017bdf,Goldberg:2019ktf,Murphy:2019eyu}. Therefore, for our purposes, we assume the energy deposition from the decays of sterile neutrinos inside the SN envelope to be less than $E_\text{dep}<10^{50}$ erg. Note that this energy deposition should occur beyond the radius of the SN core ($R_{\text{core}}$) but inside the envelope of the exploding star ($R_{\text{env}}$). 

Previous works in the literature often employ the Raffelt criterion to set a cooling bound. Our results are in agreement with these  when matching their assumptions, i.e., proton Primakoff scattering being subleading. We focus instead more on the IL bound.  There are several advantages to the IL criterion. Firstly, it is more consistent with the physical picture of the process, i.e., sterile neutrino production occurs at different times and at different radii throughout the proto-neutron star core. Secondly, as we will show later, the Raffelt criterion is not reliable to obtain bounds in the trapping regime. Since it assumes the sterile neutrino production at a specified radius, the absorption rate might be dominated by other modes apart from decays. It will be demonstrated later using IL criterion that the bounds in the trapping regime are set by the sterile neutrino decay rather than scatterings. Hence for heavy sterile neutrinos which can decay, the IL criterion is more apt. 

For our purposes, we assume $N$ production through very small transition magnetic moments do not appreciably affect the standard SN processes. In this work, we apply our reasoning to obtain bounds in the dipole coupling---mass plane with the SFHo-18.8 model simulated by the Garching group, which adopts a $18.8\, M_\odot$ progenitor and includes six-species neutrino transport \cite{ SNprofile,Mirizzi:2015eza,Bollig:2020xdr}. We use the simulated SN evolution assuming $R_\text{core}\sim 20$ km for all post-bounce time sequences up to $\sim 10$ s and assume an envelope extending up to $\sim 5 \times 10^{8}$ km.

\subsection{Absorption Modes} The decay and scatterings of $N$ can lead to novel energy deposition in the SN envelope, which can contribute to the SN explosion. The relevant processes that determine the mean free path are,  
\begin{align}
    N + e^\pm  &\rightarrow \nu + e^\pm, &\qquad\text{(downscattering)} \\
    N + \mu^\pm  &\rightarrow \nu + \mu^\pm, &\qquad\text{(downscattering)} \\
    N + p  &\rightarrow \nu + p, &\qquad\text{(downscattering)} \\
    N + \bar{\nu}  &\rightarrow \gamma, &\qquad\text{(annihilation)} \\
    N &\rightarrow \nu + \gamma. &\qquad\text{(decay)} 
\end{align}
In the absence of scatterings, the decay rate $\Gamma$ is dominated by the $N \rightarrow \nu + \gamma$ process, for which the vacuum decay rate is given by,
\begin{equation}
    \Gamma_{N \rightarrow \nu + \gamma} = \frac{d^2 M_N^3}{4\pi}.
\end{equation}
The decay length $\lambda_{\text{decay}}$ can be calculated by taking the Lorentz factor $\gamma = 1/\sqrt{1-\beta^2}$ into account, i.e., $\lambda_{\text{decay}}=\gamma\beta/\Gamma_{N \rightarrow \nu + \gamma}$, where $\beta=p_{N}/E_N$. Due to the significant population of photons and neutrinos inside the SN core, the decay rate for radiative decay will be modified. This difference occurs because of Pauli blocking of neutrinos and stimulated emission of the photon (bose enhancement) in the final state. The mean free path calculation including these effects will be described in detail later. 

Note that similar to our work in~\cite{Chauhan:2023sci}, we assume that a major portion of the outgoing energy in scattering and decay processes is carried by non-neutrino species, which are readily absorbed by the SN medium. We also point out that high energy neutrinos are most likely to be deposited. Hence, it is a good assumption that entire energy of the downscattered or decayed $N$ is deposited inside the SN.

\begin{figure*}[htbp]
    \centering
    \includegraphics[width=1.5\columnwidth]{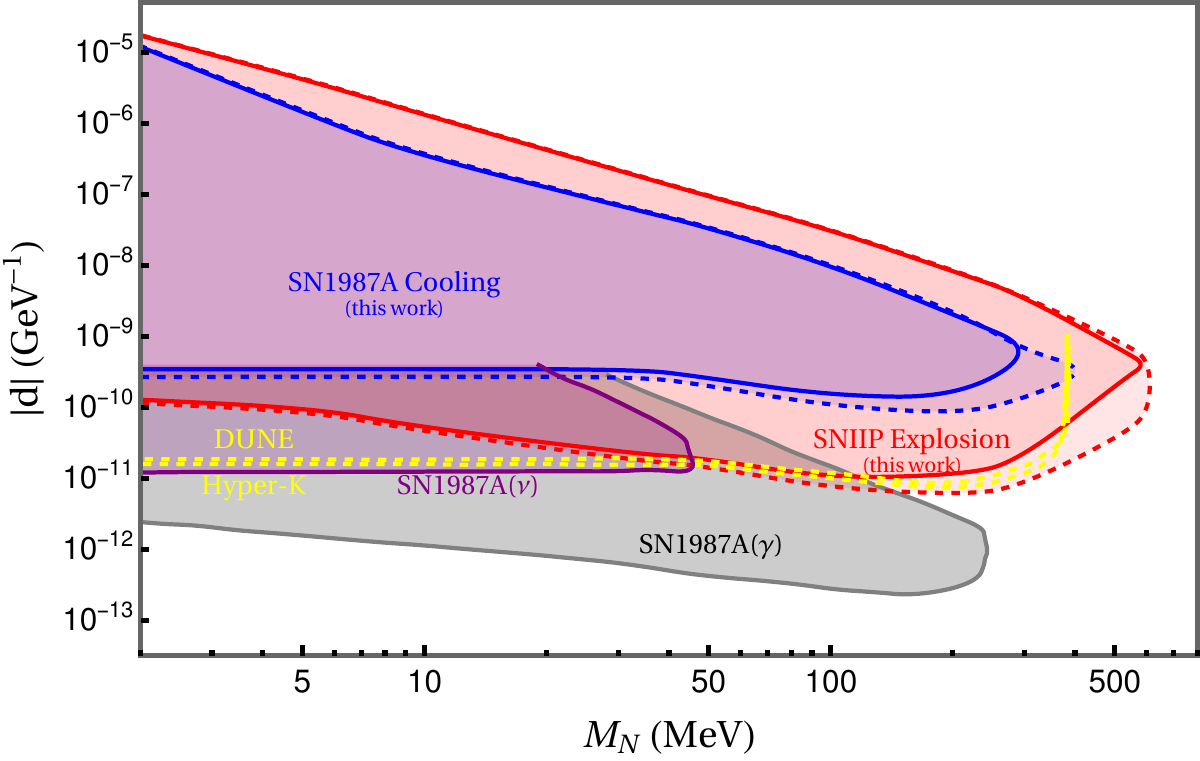}
    \caption{Bounds on flavor universal dipole strength $|d|$ from SN1987A cooling (blue) and SNIIP explosion energies (red) as functions of sterile neutrino mass, In dashed lines we show the constraints ignoring gravitational trapping, which has a noticeable effect especially at higher sterile neutrino masses. Other displayed constraints include SN1987A($\gamma$)~\cite{Brdar:2023tmi},   SN1987A($\nu$)~\cite{Brdar:2023tmi} and the  experimental sensitivity of DUNE and Hyper-K  for a future Galactic SN at a distance of $10$ kpc~\cite{Brdar:2023tmi}.}
    \label{fig:p2}
\end{figure*}

\subsection{Energy Cooling/Deposition} Our constraints arise from the sterile neutrino production in the SN core through the magnetic moment portal, with the bounds on the energy loss or deposition arising from observations of SN1987A and low-energy SNIIP, respectively. The salient details of the production processes have been discussed in previous sections. The total energy deposited or taken away ($E_{\rm dep/cool}$) from the SN core can be calculated by time-integrating the differential sterile neutrino luminosity $L_N$ over the core volume, weighted by the escape probability $P_{\text{cool/dep}}$,
\begin{widetext}
\begin{align}
    E_{\text{dep/cool}}= \eta_\text{lapse}^2 \int dt \int_0^{R_\text{core}} dr \int_{M_N}^{\infty} dE_N \frac{dL_N(r,E_N,t)}{dr\,dE_N}\Theta\left(E_N - \frac{M_N}{\eta_\text{lapse}}\right) \times P_{\text{cool/dep}}(r) ,
    \label{eq:edep}
\end{align}
\end{widetext}
where  $\eta_\text{lapse}$ is the gravitational redshift factor, $E_N$ is the sterile neutrino energy, $\frac{dL_N(r,E_N,t)}{dr\,dE_N}$ is the gradient of the differential sterile neutrino luminosity, $\Theta(x)$ is the Heaviside theta function and $P_{\text{cool/dep}}(r)$ is the probability for $N$ produced at $r$ to escape. $P_{\text{cool/dep}}$ is determined by the mean free path of the sterile neutrino in the hot dense environment of the SN. $P_{\text{cool/dep}}$ incorporates the effect of the decays and scattering of the sterile neutrino with the medium, which might prohibit the efficient transport of the energy from the core to mantle and/or beyond. Using the absorptive width of the sterile neutrino $\Gamma_{abs}$, we can define $P_{\text{cool}}$ in terms of the optical depth $\tau$~\cite{Chang:2016ntp}, 
\begin{equation}
     P_{\text{cool}}(r) = \text{exp}\left[-\tau(r,R)\right]  = \text{exp}\left[-\int_r^{R} \Gamma_{abs}(r')\, \text{d}r'\right].
    \label{eq:optD}
\end{equation}
The absorption rate for $2\rightarrow 2$ scatterings is given by an expression similar to the collisional term~\cite{Weldon:1983jn,Chang:2016ntp,Lucente:2022vuo},
\begin{align}
     \Gamma_{abs} = &\frac{1}{2 p_N} \int d^3 \Tilde{p_2} d^3 \Tilde{p_3} d^3 \Tilde{p_4}\, \Tilde{\Lambda}(f_2,f_3,f_4) \times \nonumber \\ &  |M|^2_{12\rightarrow 34}\, \delta^4(p_N+p_2-p_3-p_4) (2 \pi)^4,
\end{align}
where $\Tilde{\Lambda}(f_2,f_3,f_4)= f_2(1-f_3)(1-f_4)$.

The cooling bound is applicable only if the energy from the core can be transferred efficiently beyond the shock, where this energy cannot be reprocessed for neutrino production/streaming. 
For example, $N$ might decay before the shock radius, which will not lead to an energy loss and the cooling bound will not apply. The average probability for the energy transport beyond the neutrinosphere is given by,
\begin{equation}
    P^{\text{SN1987A}}_{\text{cool}}(r) = \text{exp}\left[-\int_r^{R_{\rm far}} \Gamma_{abs}(r')\, \text{d}r'\right]. 
    \label{eq:Pesc1}
\end{equation}
Note that we assume radial outward propagation for the calculation of the absorptive width. $R_{\rm far}$ can be defined in two different ways with the only strict requirement being $R_{\rm far}>R_{\nu}$. Usually $R_{\rm far}$ is not set very close to $R_\nu$, since the production rate from the outermost thin shell centred at $R_{\nu}$ might be overestimated. Note that the actual position of $R_{\rm far}$ is inconsequential for the bounds derived in our work as long as it is beyond the neutrinosphere, since the optical depth is dominated by the absorptive width of the high temperature region surrounding the radius of the production especially the regions just beyond $R_{\rm core}$ if the final state in the decays or scatterings is Pauli-blocked inside the core. In literature, either gain radius $R_{\rm gain}\sim \mathcal{O}(100)$ km or $R_{\rm shock}\sim \mathcal{O}(1000)$ km  is usually chosen as representative values for $R_{\rm far}$ \cite{Chang:2016ntp}. In this work, we set $R_{\rm far}$ to $R_{\rm gain}$.

For the case of low-energy SN, the bounds apply only if energy deposition takes place between $R_{\rm core}$ and the outermost envelope radius, $R_{\rm env}$. Therefore, the escape probability in this case can be written,
\begin{widetext}
    \begin{equation}
    P^{\text{SNIIP}}_{\text{dep}}(r) = \text{exp}\left[-\int_r^{R_{\rm core}} \Gamma_{abs}(r')\, \text{d}r'\right] \left( 1- \text{exp}\left[-\int_{R_{\rm core}}^{R_{\rm env}} \Gamma_{abs}(r')\, \text{d}r'\right] \right).
    \label{eq:Pesc2}
\end{equation}
\end{widetext}
For our purposes, $R_{\rm core}$ can be defined as the radius of the neutrinosphere beyond which neutrinos free stream, broadly defined as the radius at which $T_{SN}$ falls down to $3$ MeV. The actual neutrinosphere radius depends on the neutrino flavor, but assuming the same $R_{\nu}$ for all species will not affect the bound appreciably. In this work, $R_{\rm core}=20$ km and $R_{\rm env}$ is chosen to be the progenitor radius equal to $5 \times 10^{13} $ cm.

We also include the effect of gravitational trapping. In the absence of sufficient kinetic energy, the presence of high matter densities can lead to sterile neutrino getting trapped. Therefore, it is required that $E_N>m_s/\eta_\text{lapse}$, where $\eta_\text{lapse}$ relates the energy measured in the SN frame to the energy measured by an observer at infinity. We also need to account for gravitational time dilation which corrects for time interval measured locally compared to an observer at infinity. Therefore, a factor of $\eta_\text{lapse}$ for $L_N$ and another factor for the time interval $dt$ leads to the pre-factor $\eta_\text{lapse}^2$ in Eq.~\eqref{eq:edep}.

\section{Results and Discussion} 
\label{sec:results}
We display our main results in Fig.~\ref{fig:p2} for flavor universal active-sterile magnetic moment as a function of sterile neutrino mass $M_N$. The curves shown in blue are obtained through the SN1987A cooling bound $E_{N,{\rm cool}} < 10^{52}\,  {\rm erg}$. The curves shown in red are obtained through the bound on explosion energies using SNIIP, $E_{N,{\rm dep}} < 10^{50}\,  {\rm erg}$. The dashed blue and red curves are bounds from cooling and explosion energies respectively, but without taking the effect of gravitational trapping into account. The bound from SNIIP's is almost an order of magnitude stronger than the cooling bound. It can reach $|d|$ as low as $10^{-11} \text{ GeV}^{-1}$ and provides one of the leading constraint for $30 \text{ MeV }\leq M_N \leq 600 \text{ MeV}$. We also include other constraints on $|d|$ from the radiative decay of $N$ from SN1987A~\cite{Brdar:2023tmi} (labeled SN1987A($\gamma$)) and from the bound on the neutrino flux arising from radiative decay~\cite{Brdar:2023tmi} (labeled SN1987A($\nu$)). The dotted yellow curve shows the experimental sensitivity of upcoming neutrino experiments DUNE and Hyper-Kamiokande for a future galactic SN, assuming a hypothetical distance of $D_{SN}=10$ kpc~\cite{Brdar:2023tmi}. 

We observe for the SN1987A cooling bound that gravitational effects lead to trapping of $M_N \gtrsim 300$ MeV, while for the SNIIP explosion bound, although the $M_N$ range is not affected appreciably, the bounds for higher $M_N$ becomes weaker. This occurs due to gravitational trapping leading to a suppression of production rates, which can only be countered through increased coupling strength for the cooling or explosion energy bound to apply. However, increased $|d|$ required for the cooling case is beyond the trapping regime, therefore gravitational effects shrink the mass reach of the cooling bound.

In the bottom region of the blue and red curves, the production rate for low $M_N$ is dominated by proton Primakoff upscattering, followed by the electron upscattering (also see fig.~\ref{fig:prod}(a)). The production rate for Primakoff upscattering for low $M_N$ is largely independent of $M_N$ as also indicated by the flat region in the cooling bound curve. Upon increasing $M_N$, the inverse decay starts to dominate the production rate, especially above $50$ MeV. The inverse decay production rate depends on $M_N^4$ and remains dominant up to the kinematic threshold of $\sim \mu_{\nu_e} + T$ but suffers Boltzmann suppression above these masses. Since the couplings are extremely low in this regime, the exponential factor $P_{\text{esc}}\simeq 1$ in this region. 

\begin{figure}[t]
    \centering
    \includegraphics[width=\columnwidth]{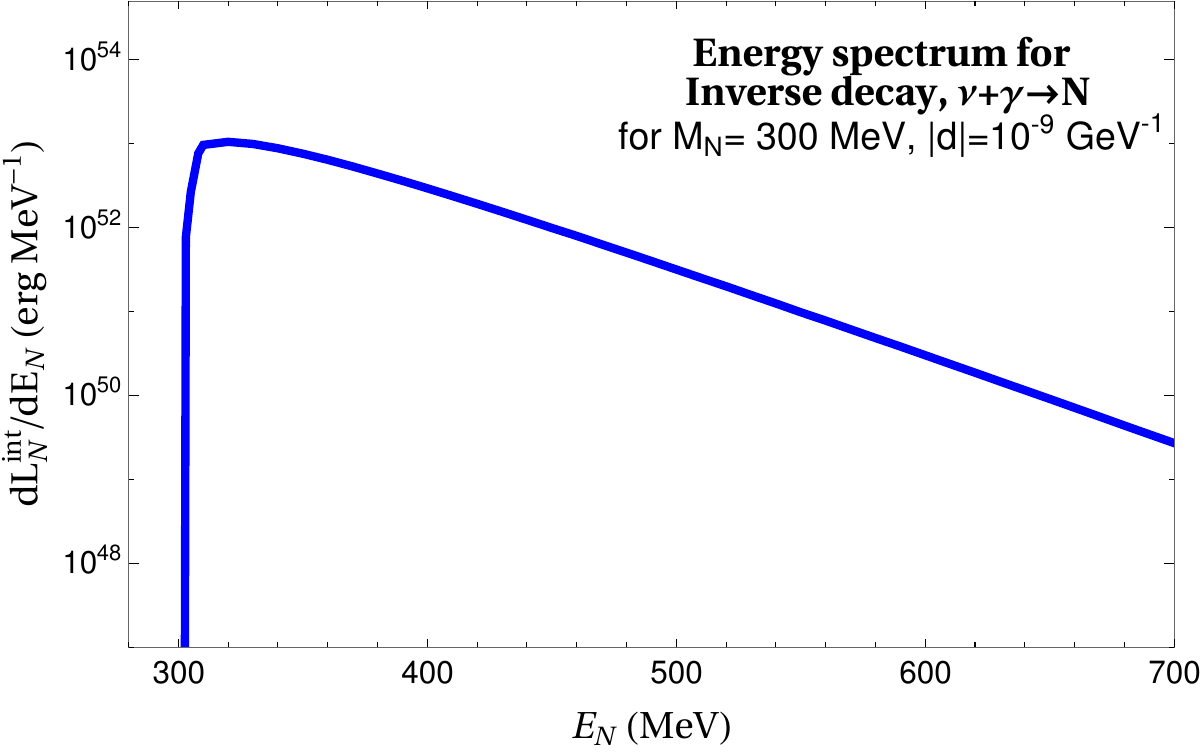}
    \caption{Differential time-integrated luminosity as a function of sterile neutrino energy for $M_N=300$ MeV and $|d|=10^{-9} \text{ GeV}^{-1}$ for inverse decay process.}
    \label{fig:energySpectrum}
\end{figure}

For the trapping regime, the coupling is set by the requirement that the mean free path length is less than $\mathcal{O}(R_{\rm core})$. In this region, the couplings are really high and therefore production regions with higher absorption rates get suppressed in the energy integral Eq.~(\eqref{eq:edep}). Therefore, the dominant contribution arises from regions with the least absorption rates, i.e., regions near $R_{\rm core}$. In these outer regions near the core, the proton and electron number density is comparatively lower, therefore the absorption rate is dominated by the decays of $N$, which sets the maximum allowed coupling strength in the trapping regime. This is in direct contrast to the Raffelt criterion where the opacity is calculated at a given radius only, and if chosen inside the core, the absorption rate might be dominated by other modes as also implemented by Ref.~\cite{Magill:2018jla}. They find the Primakoff upscattering contribution to the absorption rate to be dominant which leads to the flattening of the trapping bound at low $M_N$, which however is not the correct physical picture as pointed above. A brief discussion and comparison of their results with ours is presented in Appendix~\ref{sec:compLit}. Another important observation is the impact of the broadness of the sterile neutrino energy spectrum. For a broader energy spectrum (e.g., in Fig.~\ref{fig:energySpectrum}), higher energy $N$'s can be produced at a similar rate as compared to the assumed mean sterile neutrino energy (see Ref.~\cite{Magill:2018jla}). To trap these energetic $N$'s, the couplings need to be comparatively higher, which results in the trapping regime shifting to higher values. This is the primary reason why our trapping bound for higher masses assuming $R_{far}$ at $r=100$ km matches Ref.~\cite{Magill:2018jla} bound, which assumes $R_{far}$ at $r=25$ km.   

We also point out that our results are consistent with the cooling bound constraint in Ref.~\cite{Brdar:2023tmi} using $8.8 M_\odot$ progenitor. However, Ref.~\cite{Brdar:2023tmi} did not include the proton upscattering mode. In addition, the progenitor star for SN1987A is more than likely approximated by a $18.8 M_\odot$ progenitor than a $8.8 M_\odot$, the latter of which tends to have lower maximum temperatures which especially affects the thermal production of $N$ at high $M_N$ through inverse decays. 

We also note that the magnetic moment portal, although quite similar at first glance to the axion case~\cite{Caputo:2022mah} (both species with radiative couplings), differ qualitatively from each other. In the former case, the production rate is enhanced especially for lower $M_N$ from the high chemical potential of $\nu$ in the initial state, for both Primakoff upscattering and inverse decay processes, while no such enhancement is possible for the axion case, where the $\nu$ is replaced by the $\gamma$, which are thermally produced.    

\section{Conclusions}
We have revisited the SN1987A cooling bound and obtained new bounds from SNIIP explosion energies, for the dipole portal. We found that SNe can be efficient sites of sterile neutrino production via magnetic moments, and that the integrated luminosity criteria can produce stronger results than the Raffelt criterion, especially in the trapping regime. Secondly, we have found that low-energy supernovae can significantly cover previously unconstrained parameter space. 

We have included the effect of nucleon self-energies, Debye screening,  gravitational trapping as well as the effect of degeneracy on the production rates. In addition to including the plasmon decay channel, our work also includes the production modes arising from substantial muon population in the SNe core.

Future directions for this work motivates the calculation of exact thermal rates for Primakoff upscattering. {In light of proton Primakoff upscattering rate, the constraints derived from the neutrino and photon flux arising from the radiative decay of $N$ from SN1987A might become stronger~\cite{Brdar:2023tmi}. Another interesting case might occur, the $\gamma$ from low mass steriles decaying outside the SN might not be able to escape and could form a fireball, like in the case of axions~\cite{Diamond:2023scc}.} In addition, the bounds may be improved by refined calculation for the thermalization and trapping of $N$'s and including thermal masses of photons in $N$ decays.

\section*{Acknowledgements}

We are very grateful to Vedran Brdar, Ryan Plestid and Yingying Li for helpful discussions. We thank Hans-Thomas Janka and Daniel Kresse for providing the SN profiles used in this work. G.C would like to also thank Washington University physics department for the use of HPC Center facilities. 
The work of GC, SH, PH, and IS is supported by the U.S. Department of Energy under the award number DE-SC0020250 and DE-SC0020262. 
The work of SH is also supported by NSF Grant No.~AST1908960 and No.~PHY-2209420, and JSPS KAKENHI Grant Number JP22K03630 and JP23H04899. This work was supported by the World Premier International Research Center Initiative (WPI Initiative), MEXT, Japan.  

\appendix
\section{Comparison with Raffelt Criterion}\label{sec:compLit}

\begin{figure}[t]
    \centering
    \includegraphics[width=\columnwidth]{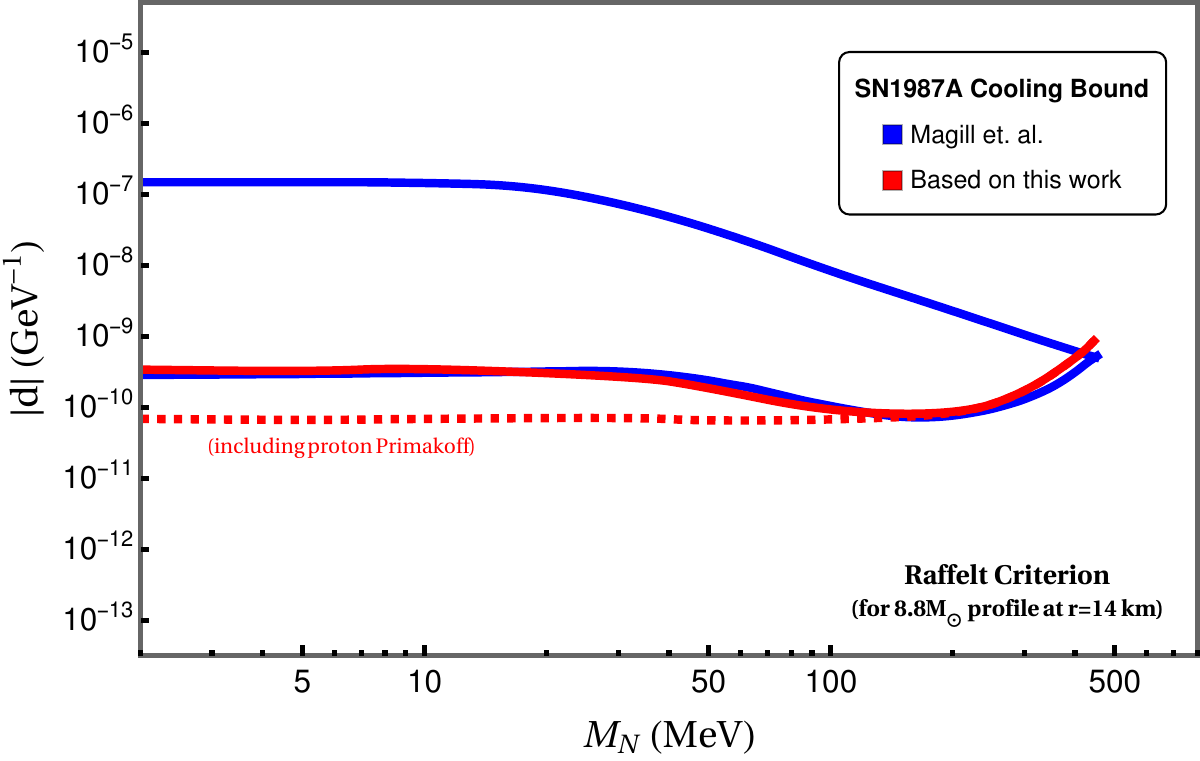}
    \caption{SN1987A cooling bound obtained using the Raffelt criterion in Refs.\cite{Magill:2018jla} (blue curve) and using the production rate calculation used in this work (red curve, for details refer to text). The dashed red line shows bounds using the Raffelt criterion including proton Primakoff upscattering.}
    \label{fig:p1}
\end{figure}

We will compare and discuss the results for cooling bound for the dipole portal obtained in Ref.~\cite{Magill:2018jla} with our results using the Raffelt criterion for the same $18.8 M_\odot$ progenitor~\cite{Fischer:2016cyd} at $r=14$ km, shown in Fig.~\ref{fig:p1}. As detailed in Ref.~\cite{Magill:2018jla}, their cooling bound (blue curve) is dominated by the electron Primakoff upscattering for lower $M_N$ and by inverse decays for higher $M_N$. For faithful comparison, we show our results for cooling bound excluding the proton upscattering mode, shown as solid red curve. It can be clearly seen that our results are in complete agreement, by excluding the upscattering off of proton. However upon including the proton upscattering process, the cooling bound becomes stronger as shown in dashed red curve. Therefore, we observe that the proton mode can help improve the constraint on the dipole portal.

As for the cooling bound in the trapping regime (as discussed earlier
in Sec.~\ref{sec:results}) since the Raffelt criterion is done at a specified radius, usually at $r<{R_{core}}$, the opacity calculation to obtain bounds does not capture the real picture. Assuming a mono-energetic sterile neutrino for trapping also affects the analysis. It only becomes clear in the implementation of integrated luminosity criterion that production rate at very high couplings inspite of the high absorption rate can still proceed from the edges of the core, therefore cooling/trapping bound is still applicable. Since at $r\sim R_{core}$, the dominant channel for energy loss/deposition is the sterile neutrino decay. Therefore, $N$ decays set the trapping regime for all $M_N$ irrespective of the other scattering modes.

\section{Collisional Integral for $s$-channel processes}\label{sec:appendixCollIntS}
For $s$-channel processes, the standard reduction of 9-dimensional collisional term to a 3-dimensional integral as detailed in Ref.~\cite{Hannestad:1995rs} fails. This happens due to the momentum transfer $q^2$ in the denominator for the matrix element of the s-channel process being a function of $\cos{\alpha}$. Due to which the usual step involving analytical integration of $\cos{\alpha}$ does not work. In this Appendix, we show how the integrals in Eq.~(\ref{eq:collint}) can be reduced from nine to three dimensions for a s-channel process. Our procedure closely follows the techniques used in Ref.~\cite{Hannestad:1995rs}. Our procedure primarily relies on swapping out the angular coordinates for $\mathbf{p_2}$ and $\mathbf{p_3}$ compared to the standard way. Note that this simple change leads to non-trivial sign and variable changes throughout the standard calculation, therefore we reproduce our entire calculation here.
We begin by using the following property, 
\begin{equation}
\frac{{\rm d}^3p_4}{2E_4}={\rm d}^4p_4\delta (p_4^2-m_4^2)\Theta(p_4^0)\,\label{deltaprop}
\end{equation} 
The integral over $p_4$ is done using the four-dimensional delta function arising from momentum conservation in the scattering process, enforcing $p_4=p_1+p_2-p_3$ throughout rest of the calculation.
We now introduce the following spherical coordinates for the 3-momenta,
\begin{align}
    \mathbf{p_1} &= p_1 (0,0,1), \\
    \mathbf{p_2} &= p_2 (0,\sin{\theta},\cos{\theta}), \\
    \mathbf{p_3} &= p_3 (\sin{\alpha}\sin{\beta},\sin{\alpha}\cos{\beta},\cos{\alpha}). 
\end{align}
The volume element for $p_2$ and $p_3$ can be written as
\begin{align}
{\rm d}^3p_2&=p_2^2{\rm d}p_2{\rm d}\cos\theta {\rm d}\mu \,\ , \\
{\rm d}^3p_3&=p_3^2{\rm d}p_3{\rm d}\cos\alpha{\rm d}\beta \,\ ,
\end{align}
with $\mu$ and $\beta$ being the azimuthal angles for $\mathbf{p_2}$ and $\mathbf{p_3}$.
The integration over $d\beta$ is carried out using $\delta(p_4^2-m_4^2)\equiv\delta(f(\beta))$, by using the relation
\begin{equation}
\int {\rm d}\beta\delta(f(\beta))=\sum_i\int {\rm d}\beta \frac{1}{\big|f'(\beta)\big|_{\beta=\beta_i}}\delta(\beta-\beta_i) \,\ ,
\end{equation}
where the $\beta_i$ are the roots of $f(\beta)=0$ and
\begin{equation}
f'(\beta)=\frac{{\rm d}f(\beta)}{{\rm d}\beta}=-2p_2p_3\sin\alpha\sin\theta\sin\beta \,\ ,
\end{equation}
with $\sin\beta_i = \pm(1-\cos^2\beta_i)^{1/2}$, where
\begin{widetext}
\begin{equation}
\cos\beta_i=\frac{2E_2E_3-2p_2p_3\cos\alpha\cos\theta-Q-2E_1E_2+2p_1p_2\cos\theta+2E_1E_3-2p_1p_3\cos\alpha}{2p_2p_3\sin\alpha\sin\theta} 
\end{equation}
\end{widetext}
and $Q\equiv m_1^2+m_2^2+m_3^2-m_4^2$. To account for the two different solutions for $\cos\beta$, we can restrict the integration interval to $[0,\pi]$ and multiply with a factor of 2. Note that since the integrand is independent of $\mu$, the integration over ${\rm d}\mu$ is trivial and equals $2\pi$.

The limits of integration in ${\rm d}\cos\alpha$ come from demanding that $\cos^2\beta\leq 1$. This requirement can also be stated as 
\begin{equation}
(f'(\beta))^2=(2p_2p_3\sin\alpha\sin\theta\sin\beta)^2\geq 0 
\end{equation}
Therefore we can write
\begin{equation}
\int_0^{2\pi}{\rm d}\beta\delta(f(\beta))=2\frac{1}{|f'(\beta)|_{\beta=\beta_i}}\Theta\Bigg(|f'(\beta)|_{\beta=\beta_i}^2\Bigg) \,\ .
\end{equation}
To simplify the expressions, we introduce the following definitions:
\begin{align*}
\gamma&=E_1E_2-E_1E_3-E_2E_3 \,\ ; \\
\epsilon&=-p_1p_2\cos\theta \,\ ; \\
k&=p_1^2+p_2^2 \,\ ; \\
a&=p_3^2(-4k+8\epsilon) \,\ ; \\
b&=-p_3(p_1-\epsilon/p_1)(8\gamma+4Q+8\epsilon) \,\ ; \\
c&=-4\gamma^2-4\gamma Q-Q^2-8\gamma\epsilon-4Q\epsilon-4\epsilon^2 \\ &+4p_2^2p_3^2(1-\cos^2\theta) \,\ ;
\end{align*}
With the above notation, $f'(\beta)$ can be written as:
\begin{equation}
|f'(\beta)|_{\beta=\beta_i}=\sqrt{a\cos^2\alpha+b\cos\alpha+c} \,
\end{equation}
All possible matrix elements only include products of the four-momenta, which are calculated below:
\begin{align*}
p_1\cdot p_2&=E_1E_2-p_1p_2\cos\theta \,\ , \\
p_1\cdot p_3&=E_1E_3-p_1p_3\cos\alpha \,\ , \\
p_1\cdot p_4&=m_1^2+(E_1E_2-p_1p_2\cos\theta)-(E_1E_3-p_1p_3\cos\alpha) \,\ , \\
p_2\cdot p_3&= (E_1E_2-p_1p_2\cos\theta)-(E_1E_3-p_1p_3\cos\alpha)+\frac{Q}{2} \,\ , \\
p_2\cdot p_4&=(E_1E_3-p_1p_3\cos\alpha)+m_2^2-\frac{Q}{2} \,\, \\
p_3\cdot p_4&=(E_1E_2-p_1p_2\cos\theta)-m_3^2+\frac{Q}{2} \,\ . 
\end{align*}
Now it can be checked that all $s$-channel processes are analytically integrable over ${\rm d}\cos\alpha$ and can be carried out by using these relations~\cite{Mastrototaro:2021wzl}:
\begin{widetext}
\begin{align*}
\int\frac{1}{\sqrt{ax^2+bx+c}}\Theta(ax^2+bx+c){\rm d}x&=\frac{\pi}{\sqrt{-a}}\Theta(b^2-4ac)\,\ ; \\
\int\frac{x}{\sqrt{ax^2+bx+c}}\Theta(ax^2+bx+c){\rm d}x&=-\frac{b}{2a}\frac{\pi}{\sqrt{-a}}\Theta(b^2-4ac)\,\ ; \\
\int\frac{x^2}{\sqrt{ax^2+bx+c}}\Theta(ax^2+bx+c){\rm d}x&=\Bigg(\frac{3b^2}{8a^2}-\frac{c}{2a}\Bigg)\frac{\pi}{\sqrt{-a}}\Theta(b^2-4ac)\,\ . 
\end{align*}     
\end{widetext}
The step function arises from demanding a real integration interval. This also ensures that the roots of $ax^2+bx+c$ are not outside the fundamental integration interval of $[-1,1]$. Similarly, the integration interval for integration over ${\rm d}\cos\theta$ is given by the solutions of $b^2-4ac=0$:
\begin{equation}
\cos\theta=\frac{2\gamma+2p_3^2+Q\pm2p_3\sqrt{2\gamma+p_1^2+p_2^2+p_3^2+Q}}{2p_1p_2} \,\ .
\end{equation}
For the integration interval to be real, both of these solutions are required to be real. We refer to these two solutions as $\cos\theta_{\mathrm{-}}$ and $\cos\theta_{\mathrm{+}}$. The real integration limits are $\alpha=\sup[-1,\cos\theta_{\rm -}]$ and $\beta=\inf[+1,\cos\theta_{\rm +}]$ with $\alpha\leq\beta$.
Finally by combining all the analytical simplifications described above, Eq.~\eqref{eq:collint} is reduced to the following three dimensional integral, which is evaluated numerically :
\begin{widetext}
\begin{equation}
\mathcal{C}_{coll}(f_1)=\frac{2}{(2\pi)^4}\frac{1}{2E_1}\int_0^{\infty}\int_0^{p_1+p_2}\int_{\alpha}^{\beta}{\rm d}\cos\theta\,\frac{p_2^2{\rm d}p_2}{2E_2}\frac{p_3^2{\rm d}p_3}{2E_3}\Lambda(f_1,f_2,f_3,f_4) F(p_1,p_2,p_3)\,\Theta(A) \,\ ,
 \end{equation}     
\end{widetext}
where $A$ is the parameter space allowed i.e. $\alpha,\beta \in \mathbb{R}$, $\alpha\leq\beta$ and $F$ is derived from the following analytical integral:
\begin{widetext}
\begin{equation}
F(p_1,p_2,p_3)\equiv \int\frac{|M|^2}{\sqrt{a\cos^2\alpha+b\cos\alpha+c}}\Theta(a\cos^2\alpha+b\cos\alpha+c){\rm d}\cos\alpha \,\,
\end{equation}
\end{widetext}

\section{Collisional Integral for $N \leftrightarrow \nu + \gamma$ }\label{sec:appendixInv}
The matrix element for the decay process $N \rightarrow \nu + \gamma$ is
\begin{equation}
    |\mathcal{M}|^2 = 2\,d^2 (M_N^2-m_\nu^2)^2
    \label{eq:matrixInv}
\end{equation}
The collision term for $2\rightarrow 1$ inverse decay in this case is \cite{Chang:2016ntp,Lucente:2022vuo}
\begin{equation}
    \begin{aligned}
    \mathcal{C}_{\nu + \gamma \rightarrow N} = \frac{1}{2 E_N} \int &\frac{d^3 {p_\gamma}}{(2 \pi^3)\,2 E_\gamma} \frac{d^3 {p_\nu}}{(2 \pi^3)\,2 E_\nu}\, f_\gamma(E_\gamma)\,f_\nu(E_\nu) \, \\ & \times |M|^2_{N \rightarrow \nu + \gamma}\, \delta^4(p_N-p_\gamma-p_\nu) (2 \pi)^4
    \end{aligned}
\end{equation}
where $f_i(E)$ is the respective quantum-statistics factor i.e. Bose-Einstein or Fermi-Dirac, for the initial states. The above 6-dimensional integral can be reduced to the following 1-dimensional integral
\begin{equation}
\begin{aligned}
     \mathcal{C}_{\nu + \gamma \rightarrow N} = \frac{d^2\,M_N^4}{16\pi p_N E_N} \int_{P^-}^{P^+} & d{p_\gamma} \, f_\gamma(p_\gamma) \\ &\times f_\nu\left(\sqrt{p_N^2+M_N^2}-p_\gamma\right)
\end{aligned}
\end{equation}
where $P^{\pm}=(E_N\pm p_N)/2$. 

Similarly, the absorption rate $\Gamma$ in a medium composed of photons and neutrinos can be written as
\begin{equation}
\begin{aligned}
     \Gamma_{N \rightarrow \nu + \gamma} = \frac{d^2\,M_N^4}{16\pi p_N^2} & \int_{P^-}^{P^+}  d{p_\gamma} \, (1+f_\gamma(p_\gamma)) \\ & \times \left[1-f_\nu\left(\sqrt{p_N^2+M_N^2}-p_\gamma\right)\right]
\end{aligned}
\end{equation}
In absence of a medium, the thermal distributions vanish and yield the vacuum decay rate. This difference occurs because of Pauli blocking of neutrinos i.e. $(1-f_\nu(E))$ and stimulated emission of the photon (bose enhancement) i.e. $(1+f_\gamma(E))$ in the final state. 

\section{Plasmon Decay}\label{sec:appendixPlasm}
The decay rate for $\gamma^* \rightarrow N + \Bar{\nu}$, applicable to both transverse and longitudinal excitations is given by \cite{Vogel:2013raa,Brdar:2020quo}, 
\begin{equation}
    \begin{aligned}
        \Gamma_{\gamma^*} = \frac{d^2}{24 \pi}\mathcal{Z}\frac{(\omega^2-k^2)^2}{\omega} \left(1-\frac{M_N^2}{K^2} \right)^2 \\ \times \left(1+2\frac{M_N^2}{K^2} \right) \Theta(K-M_N)
    \end{aligned}
    \label{eq:GammaPlasmon}
\end{equation}
where $\mathcal{Z}$ is the renormalization constant, $K^2=\omega^2-k^2$ is the effective plasmon mass, $\omega$ and $k$ are plasmon energy and momentum.

The total energy loss rate including contributions from both transverse and longitudinal plasmons can be written as~\cite{Raffelt:1996wa,Sutherland:1975dr} 
\begin{equation}
    Q_{\gamma^*}= 2\int_0^\infty \frac{k^2\,dk}{2 \pi^2} \frac{\omega\Gamma_{\gamma^*}^T}{e^{\omega/T}-1} + \int_0^{k_1} \frac{k^2\,dk}{2\pi^2} \frac{\omega\Gamma_{\gamma^*}^L}{e^{\omega/T}-1} 
    \label{eq:Qplasmon}
\end{equation}
where the factor of 2 stands for two polarization states of the transverse plasmon, $\Gamma_{\gamma^*}^{T,L}$ is given by Eq.~\eqref{eq:GammaPlasmon} with appropriate renormalization factors and dispersion relations. For longitudinal modes, the momentum integration is only allowed upto $k<k_1$. It is defined as the wavenumber where $\omega(k)$ crosses the light cone i.e. $\omega/k=1$, 
\begin{equation}
    k_1^2=\frac{3\omega_P^2}{v_*^2}\left[ \frac{1}{2v_*}\log{\left( \frac{1+v_*}{1-v_*} \right)}  -1 \right]
\end{equation}
where $\omega_P$ is the plasma frequency, $v^*$ is a ``typical'' electron velocity. For modes above $k_1$, the four-momentum of a longitudinal excitation becomes space-like, are kinematically forbidden to decay.

The photon dispersion relations for a general medium are given by the following transcendental equations~\cite{Braaten:1993jw}
\begin{equation}
    \begin{aligned}
        \omega^2-k^2 & =  \omega_P^2\left[1+ \frac{ 1}{2} G(v_*^2 k^2/\omega^2)\right], &\quad \text{Transverse} \\
        \omega^2- v_*^2k^2 & =  \omega_P^2\left[1- G(v_*^2 k^2/\omega^2)\right], &\quad \text{Longitudinal}
    \end{aligned}
\end{equation}
where $\omega_P$ is the plasma frequency, $v^*$ is a ``typical'' electron velocity and $G(x)$ is a function defined by 
\begin{equation}
    G(x)=\frac{3}{x}\left[ 1- \frac{2x}{3} -\frac{1-x}{2\sqrt{x}}\log{\left(\frac{1+\sqrt{x}}{1-\sqrt{x}} \right)} \right]
\end{equation}
For highly-degenerate relativistic plasmas, as in our case, 
\begin{equation}
    \begin{aligned}
        v^* & \simeq 1,\quad k_1 \simeq \infty \\
        \omega_P^2 & \simeq \frac{4\alpha}{3\pi}\left( \mu_e^2 + \frac{\pi^2 T^2}{3}\right)  \\
    \end{aligned}
\end{equation}
Let us look at some interesting limits for the dispersion relations in a SN core. At low momentum, $G(x) \simeq 0$ implying $\omega^2-k^2 =  \omega_P^2$ for both transverse and longitudinal modes. While for high momentum modes implying $G(x) \simeq 1$, the dispersion relations have the following form:
\begin{equation}
    \begin{aligned}
        \omega^2-k^2 & =  \frac{3}{2}\omega_P^2, &\quad \text{Transverse} \\
        \omega^2-k^2 & =  0, &\quad \text{Longitudinal}
    \end{aligned}
    \label{eq:DispersionSN}
\end{equation} 
Using Eqns.~\eqref{eq:DispersionSN} and \eqref{eq:GammaPlasmon}, we conclude that for high momentum modes, the decays of longitudinal photon into massive sterile neutrinos becomes kinematically forbidden for relativistic plasmas. Therefore, the main contribution from longitudinal modes arises from low-momentum modes but since the production rate depends on $k^2$, we expect this contribution to be sub-dominant to the production through the transverse modes.


The renormalization constants for both transverse and longitudinal modes in highly-degenerate relativistic plasmas are~\cite{Braaten:1993jw}
\begin{align}
    \mathcal{Z}_T &= \frac{2 \omega^2(\omega^2-k^2)}{3\omega_P^2\omega^2+(\omega^4-k^4)-2\omega^2(\omega^2-k^2)} \\
    \mathcal{Z}_L &= \frac{2 (\omega^2-k^2)}{3\omega_P^2-(\omega^2-k^2)}
\end{align}

\section{Primakoff scattering}\label{sec:appendixPrim}
The matrix element for the Primakoff upscattering process $\nu(p_3) + f(p_4) \rightarrow N(p_1) + f(p_2)$, where $f=e^\pm,\mu^\pm,p$ is
\begin{widetext}
\begin{equation}
    \begin{aligned}
    |\mathcal{M}|^2  = & \frac{4\,d^2\,e^2}{q^4}\left[8\,(p_1.p_2)(p_2.p_3)(p_1.p_2-p_2.p_3)\right.  -\left. 2\,M_N^2(p_1.p_2-p_2.p_3)(p_1.p_2+p_2.p_3+m_f^2)\right. + \left. \,M_N^4(p_1.p_2-p_2.p_3-m_f^2)\right] 
\end{aligned}
\label{eq:Mprim}
\end{equation}
\end{widetext}
where $q^2=(p_1.p_2-p_2.p_3)$. Note that for the case of proton, nucleon charge form factor needs to be taken into account. The form factor $F_1(q^2)$ can be obtained by solving the following pair of equations \cite{Beck:2001dz,Magill:2018jla}
\begin{align}
    F_1 - \frac{|q|^2}{4 m_p^2}F_2  = G_D & \\
    F_1 + F_2  = \mu_{p,\gamma} G_D &
\end{align}
where $\mu_{p,\gamma}=2.793$ and $G_D=1/(1+|q|^2/0.71\text{ GeV}^2)^2$.

For $\Bar{f}+ f \rightarrow N + \Bar{\nu}$, the matrix element can be obtained using crossing symmetry rules applied to the $|\mathcal{M}|^2$ for $\nu + f \rightarrow N + f$ given above. Since, it is a $s$-channel process, it does not suffer from singularities unlike $t$-channel processes. 

\bibliography{ref}

\providecommand{\href}[2]{#2}\begingroup\raggedright\begin{thebibliography}{10}

\bibitem{Abdullahi:2022jlv}
A.~M. Abdullahi et~al., \emph{{The present and future status of heavy neutral leptons}}, \href{https://doi.org/10.1088/1361-6471/ac98f9}{\emph{J. Phys. G} {\bfseries 50} (2023) 020501} [\href{https://arxiv.org/abs/2203.08039}{{\ttfamily 2203.08039}}].

\bibitem{Dolgov:2002wy}
A.~D. Dolgov, \emph{{Neutrinos in cosmology}}, \href{https://doi.org/10.1016/S0370-1573(02)00139-4}{\emph{Phys. Rept.} {\bfseries 370} (2002) 333} [\href{https://arxiv.org/abs/hep-ph/0202122}{{\ttfamily hep-ph/0202122}}].

\bibitem{Coloma:2017ppo}
P.~Coloma, P.~A.~N. Machado, I.~Martinez-Soler and I.~M. Shoemaker, \emph{{Double-Cascade Events from New Physics in Icecube}}, \href{https://doi.org/10.1103/PhysRevLett.119.201804}{\emph{Phys. Rev. Lett.} {\bfseries 119} (2017) 201804} [\href{https://arxiv.org/abs/1707.08573}{{\ttfamily 1707.08573}}].

\bibitem{Plestid:2020vqf}
R.~Plestid, \emph{{Luminous solar neutrinos I: Dipole portals}}, \href{https://doi.org/10.1103/PhysRevD.104.075027}{\emph{Phys. Rev. D} {\bfseries 104} (2021) 075027} [\href{https://arxiv.org/abs/2010.04193}{{\ttfamily 2010.04193}}].

\bibitem{Brdar:2020quo}
V.~Brdar, A.~Greljo, J.~Kopp and T.~Opferkuch, \emph{{The Neutrino Magnetic Moment Portal: Cosmology, Astrophysics, and Direct Detection}}, \href{https://doi.org/10.1088/1475-7516/2021/01/039}{\emph{JCAP} {\bfseries 01} (2021) 039} [\href{https://arxiv.org/abs/2007.15563}{{\ttfamily 2007.15563}}].

\bibitem{Brdar:2023tmi}
V.~Brdar, A.~de~Gouv\^ea, Y.-Y. Li and P.~A.~N. Machado, \emph{{Neutrino magnetic moment portal and supernovae: New constraints and multimessenger opportunities}}, \href{https://doi.org/10.1103/PhysRevD.107.073005}{\emph{Phys. Rev. D} {\bfseries 107} (2023) 073005} [\href{https://arxiv.org/abs/2302.10965}{{\ttfamily 2302.10965}}].

\bibitem{Gninenko:2009ks}
S.~N. Gninenko, \emph{{The MiniBooNE anomaly and heavy neutrino decay}}, \href{https://doi.org/10.1103/PhysRevLett.103.241802}{\emph{Phys. Rev. Lett.} {\bfseries 103} (2009) 241802} [\href{https://arxiv.org/abs/0902.3802}{{\ttfamily 0902.3802}}].

\bibitem{Gninenko:2010pr}
S.~N. Gninenko, \emph{{A resolution of puzzles from the LSND, KARMEN, and MiniBooNE experiments}}, \href{https://doi.org/10.1103/PhysRevD.83.015015}{\emph{Phys. Rev. D} {\bfseries 83} (2011) 015015} [\href{https://arxiv.org/abs/1009.5536}{{\ttfamily 1009.5536}}].

\bibitem{McKeen:2010rx}
D.~McKeen and M.~Pospelov, \emph{{Muon Capture Constraints on Sterile Neutrino Properties}}, \href{https://doi.org/10.1103/PhysRevD.82.113018}{\emph{Phys. Rev. D} {\bfseries 82} (2010) 113018} [\href{https://arxiv.org/abs/1011.3046}{{\ttfamily 1011.3046}}].

\bibitem{Masip:2011qb}
M.~Masip and P.~Masjuan, \emph{{Heavy-neutrino decays at neutrino telescopes}}, \href{https://doi.org/10.1103/PhysRevD.83.091301}{\emph{Phys. Rev. D} {\bfseries 83} (2011) 091301} [\href{https://arxiv.org/abs/1103.0689}{{\ttfamily 1103.0689}}].

\bibitem{Masip:2012ke}
M.~Masip, P.~Masjuan and D.~Meloni, \emph{{Heavy neutrino decays at MiniBooNE}}, \href{https://doi.org/10.1007/JHEP01(2013)106}{\emph{JHEP} {\bfseries 01} (2013) 106} [\href{https://arxiv.org/abs/1210.1519}{{\ttfamily 1210.1519}}].

\bibitem{Magill:2018jla}
G.~Magill, R.~Plestid, M.~Pospelov and Y.-D. Tsai, \emph{{Dipole Portal to Heavy Neutral Leptons}}, \href{https://doi.org/10.1103/PhysRevD.98.115015}{\emph{Phys. Rev. D} {\bfseries 98} (2018) 115015} [\href{https://arxiv.org/abs/1803.03262}{{\ttfamily 1803.03262}}].

\bibitem{Shoemaker:2018vii}
I.~M. Shoemaker and J.~Wyenberg, \emph{{Direct Detection Experiments at the Neutrino Dipole Portal Frontier}}, \href{https://doi.org/10.1103/PhysRevD.99.075010}{\emph{Phys. Rev. D} {\bfseries 99} (2019) 075010} [\href{https://arxiv.org/abs/1811.12435}{{\ttfamily 1811.12435}}].

\bibitem{Arguelles:2018mtc}
C.~A. Arg\"uelles, M.~Hostert and Y.-D. Tsai, \emph{{Testing New Physics Explanations of the MiniBooNE Anomaly at Neutrino Scattering Experiments}}, \href{https://doi.org/10.1103/PhysRevLett.123.261801}{\emph{Phys. Rev. Lett.} {\bfseries 123} (2019) 261801} [\href{https://arxiv.org/abs/1812.08768}{{\ttfamily 1812.08768}}].

\bibitem{Fischer:2019fbw}
O.~Fischer, A.~Hern\'andez-Cabezudo and T.~Schwetz, \emph{{Explaining the MiniBooNE excess by a decaying sterile neutrino with mass in the 250 MeV range}}, \href{https://doi.org/10.1103/PhysRevD.101.075045}{\emph{Phys. Rev. D} {\bfseries 101} (2020) 075045} [\href{https://arxiv.org/abs/1909.09561}{{\ttfamily 1909.09561}}].

\bibitem{Coloma:2019htx}
P.~Coloma, P.~Hern\'andez, V.~Mu\~noz and I.~M. Shoemaker, \emph{{New constraints on Heavy Neutral Leptons from Super-Kamiokande data}}, \href{https://doi.org/10.1140/epjc/s10052-020-7795-z}{\emph{Eur. Phys. J. C} {\bfseries 80} (2020) 235} [\href{https://arxiv.org/abs/1911.09129}{{\ttfamily 1911.09129}}].

\bibitem{Schwetz:2020xra}
T.~Schwetz, A.~Zhou and J.-Y. Zhu, \emph{{Constraining active-sterile neutrino transition magnetic moments at DUNE near and far detectors}}, \href{https://doi.org/10.1007/JHEP07(2021)200}{\emph{JHEP} {\bfseries 21} (2020) 200} [\href{https://arxiv.org/abs/2105.09699}{{\ttfamily 2105.09699}}].

\bibitem{Arina:2020mxo}
C.~Arina, A.~Cheek, K.~Mimasu and L.~Pagani, \emph{{Light and Darkness: consistently coupling dark matter to photons via effective operators}}, \href{https://doi.org/10.1140/epjc/s10052-021-09010-1}{\emph{Eur. Phys. J. C} {\bfseries 81} (2021) 223} [\href{https://arxiv.org/abs/2005.12789}{{\ttfamily 2005.12789}}].

\bibitem{Shoemaker:2020kji}
I.~M. Shoemaker, Y.-D. Tsai and J.~Wyenberg, \emph{{Active-to-sterile neutrino dipole portal and the XENON1T excess}}, \href{https://doi.org/10.1103/PhysRevD.104.115026}{\emph{Phys. Rev. D} {\bfseries 104} (2021) 115026} [\href{https://arxiv.org/abs/2007.05513}{{\ttfamily 2007.05513}}].

\bibitem{Abdullahi:2020nyr}
A.~Abdullahi, M.~Hostert and S.~Pascoli, \emph{{A dark seesaw solution to low energy anomalies: MiniBooNE, the muon (g\ensuremath{-}2), and BaBar}}, \href{https://doi.org/10.1016/j.physletb.2021.136531}{\emph{Phys. Lett. B} {\bfseries 820} (2021) 136531} [\href{https://arxiv.org/abs/2007.11813}{{\ttfamily 2007.11813}}].

\bibitem{Shakeri:2020wvk}
S.~Shakeri, F.~Hajkarim and S.-S. Xue, \emph{{Shedding New Light on Sterile Neutrinos from XENON1T Experiment}}, \href{https://doi.org/10.1007/JHEP12(2020)194}{\emph{JHEP} {\bfseries 12} (2020) 194} [\href{https://arxiv.org/abs/2008.05029}{{\ttfamily 2008.05029}}].

\bibitem{Atkinson:2021rnp}
M.~Atkinson, P.~Coloma, I.~Martinez-Soler, N.~Rocco and I.~M. Shoemaker, \emph{{Heavy Neutrino Searches through Double-Bang Events at Super-Kamiokande, DUNE, and Hyper-Kamiokande}}, \href{https://doi.org/10.1007/JHEP04(2022)174}{\emph{JHEP} {\bfseries 04} (2022) 174} [\href{https://arxiv.org/abs/2105.09357}{{\ttfamily 2105.09357}}].

\bibitem{Cho:2021yxk}
W.~Cho, K.-Y. Choi and O.~Seto, \emph{{Sterile neutrino dark matter with dipole interaction}}, \href{https://doi.org/10.1103/PhysRevD.105.015016}{\emph{Phys. Rev. D} {\bfseries 105} (2022) 015016} [\href{https://arxiv.org/abs/2108.07569}{{\ttfamily 2108.07569}}].

\bibitem{Dasgupta:2021fpn}
A.~Dasgupta, S.~K. Kang and J.~E. Kim, \emph{{Probing neutrino dipole portal at COHERENT experiment}}, \href{https://doi.org/10.1007/JHEP11(2021)120}{\emph{JHEP} {\bfseries 11} (2021) 120} [\href{https://arxiv.org/abs/2108.12998}{{\ttfamily 2108.12998}}].

\bibitem{Arguelles:2021dqn}
C.~A. Arg\"uelles, N.~Foppiani and M.~Hostert, \emph{{Heavy neutral leptons below the kaon mass at hodoscopic neutrino detectors}}, \href{https://doi.org/10.1103/PhysRevD.105.095006}{\emph{Phys. Rev. D} {\bfseries 105} (2022) 095006} [\href{https://arxiv.org/abs/2109.03831}{{\ttfamily 2109.03831}}].

\bibitem{Ismail:2021dyp}
A.~Ismail, S.~Jana and R.~M. Abraham, \emph{{Neutrino up-scattering via the dipole portal at forward LHC detectors}}, \href{https://doi.org/10.1103/PhysRevD.105.055008}{\emph{Phys. Rev. D} {\bfseries 105} (2022) 055008} [\href{https://arxiv.org/abs/2109.05032}{{\ttfamily 2109.05032}}].

\bibitem{Miranda:2021kre}
O.~G. Miranda, D.~K. Papoulias, O.~Sanders, M.~T\'ortola and J.~W.~F. Valle, \emph{{Low-energy probes of sterile neutrino transition magnetic moments}}, \href{https://doi.org/10.1007/JHEP12(2021)191}{\emph{JHEP} {\bfseries 12} (2021) 191} [\href{https://arxiv.org/abs/2109.09545}{{\ttfamily 2109.09545}}].

\bibitem{Bolton:2021pey}
P.~D. Bolton, F.~F. Deppisch, K.~Fridell, J.~Harz, C.~Hati and S.~Kulkarni, \emph{{Probing active-sterile neutrino transition magnetic moments with photon emission from CE\ensuremath{\nu}NS}}, \href{https://doi.org/10.1103/PhysRevD.106.035036}{\emph{Phys. Rev. D} {\bfseries 106} (2022) 035036} [\href{https://arxiv.org/abs/2110.02233}{{\ttfamily 2110.02233}}].

\bibitem{Jodlowski:2020vhr}
K.~Jod\l{}owski and S.~Trojanowski, \emph{{Neutrino beam-dump experiment with FASER at the LHC}}, \href{https://doi.org/10.1007/JHEP05(2021)191}{\emph{JHEP} {\bfseries 05} (2021) 191} [\href{https://arxiv.org/abs/2011.04751}{{\ttfamily 2011.04751}}].

\bibitem{Vergani:2021tgc}
S.~Vergani, N.~W. Kamp, A.~Diaz, C.~A. Arg\"uelles, J.~M. Conrad, M.~H. Shaevitz et~al., \emph{{Explaining the MiniBooNE excess through a mixed model of neutrino oscillation and decay}}, \href{https://doi.org/10.1103/PhysRevD.104.095005}{\emph{Phys. Rev. D} {\bfseries 104} (2021) 095005} [\href{https://arxiv.org/abs/2105.06470}{{\ttfamily 2105.06470}}].

\bibitem{Pauli:1930pc}
W.~Pauli, \emph{{Dear radioactive ladies and gentlemen}}, {\emph{Phys. Today} {\bfseries 31N9} (1978) 27}.

\bibitem{Caputo:2022mah}
A.~Caputo, H.-T. Janka, G.~Raffelt and E.~Vitagliano, \emph{{Low-Energy Supernovae Severely Constrain Radiative Particle Decays}}, \href{https://doi.org/10.1103/PhysRevLett.128.221103}{\emph{Phys. Rev. Lett.} {\bfseries 128} (2022) 221103} [\href{https://arxiv.org/abs/2201.09890}{{\ttfamily 2201.09890}}].

\bibitem{Chauhan:2023sci}
G.~Chauhan, S.~Horiuchi, P.~Huber and I.~M. Shoemaker, \emph{{Low-Energy Supernovae Bounds on Sterile Neutrinos}},  \href{https://arxiv.org/abs/2309.05860}{{\ttfamily 2309.05860}}.

\bibitem{Aparici:2009fh}
A.~Aparici, K.~Kim, A.~Santamaria and J.~Wudka, \emph{{Right-handed neutrino magnetic moments}}, \href{https://doi.org/10.1103/PhysRevD.80.013010}{\emph{Phys. Rev. D} {\bfseries 80} (2009) 013010} [\href{https://arxiv.org/abs/0904.3244}{{\ttfamily 0904.3244}}].

\bibitem{Babu:2020ivd}
K.~S. Babu, S.~Jana and M.~Lindner, \emph{{Large Neutrino Magnetic Moments in the Light of Recent Experiments}}, \href{https://doi.org/10.1007/JHEP10(2020)040}{\emph{JHEP} {\bfseries 10} (2020) 040} [\href{https://arxiv.org/abs/2007.04291}{{\ttfamily 2007.04291}}].

\bibitem{Raffelt:1996wa}
G.~G. Raffelt, \emph{{Stars as laboratories for fundamental physics}}. University of Chicago Press, 1996.

\bibitem{Mastrototaro:2019vug}
L.~Mastrototaro, A.~Mirizzi, P.~D. Serpico and A.~Esmaili, \emph{{Heavy sterile neutrino emission in core-collapse supernovae: Constraints and signatures}}, \href{https://doi.org/10.1088/1475-7516/2020/01/010}{\emph{JCAP} {\bfseries 01} (2020) 010} [\href{https://arxiv.org/abs/1910.10249}{{\ttfamily 1910.10249}}].

\bibitem{Hannestad:1995rs}
S.~Hannestad and J.~Madsen, \emph{{Neutrino decoupling in the early universe}}, \href{https://doi.org/10.1103/PhysRevD.52.1764}{\emph{Phys. Rev. D} {\bfseries 52} (1995) 1764} [\href{https://arxiv.org/abs/astro-ph/9506015}{{\ttfamily astro-ph/9506015}}].

\bibitem{Mastrototaro:2021wzl}
L.~Mastrototaro, P.~D. Serpico, A.~Mirizzi and N.~Saviano, \emph{{Massive sterile neutrinos in the early Universe: From thermal decoupling to cosmological constraints}}, \href{https://doi.org/10.1103/PhysRevD.104.016026}{\emph{Phys. Rev. D} {\bfseries 104} (2021) 016026} [\href{https://arxiv.org/abs/2104.11752}{{\ttfamily 2104.11752}}].

\bibitem{Hahn-Woernle:2009jyb}
F.~Hahn-Woernle, M.~Plumacher and Y.~Y.~Y. Wong, \emph{{Full Boltzmann equations for leptogenesis including scattering}}, \href{https://doi.org/10.1088/1475-7516/2009/08/028}{\emph{JCAP} {\bfseries 08} (2009) 028} [\href{https://arxiv.org/abs/0907.0205}{{\ttfamily 0907.0205}}].

\bibitem{Tamborra:2017ubu}
I.~Tamborra, L.~Huedepohl, G.~Raffelt and H.-T. Janka, \emph{{Flavor-dependent neutrino angular distribution in core-collapse supernovae}}, \href{https://doi.org/10.3847/1538-4357/aa6a18}{\emph{Astrophys. J.} {\bfseries 839} (2017) 132} [\href{https://arxiv.org/abs/1702.00060}{{\ttfamily 1702.00060}}].

\bibitem{Martinez-Pinedo:2012eaj}
G.~Martinez-Pinedo, T.~Fischer, A.~Lohs and L.~Huther, \emph{{Charged-current weak interaction processes in hot and dense matter and its impact on the spectra of neutrinos emitted from proto-neutron star cooling}}, \href{https://doi.org/10.1103/PhysRevLett.109.251104}{\emph{Phys. Rev. Lett.} {\bfseries 109} (2012) 251104} [\href{https://arxiv.org/abs/1205.2793}{{\ttfamily 1205.2793}}].

\bibitem{Mirizzi:2015eza}
A.~Mirizzi, I.~Tamborra, H.-T. Janka, N.~Saviano, K.~Scholberg, R.~Bollig et~al., \emph{{Supernova Neutrinos: Production, Oscillations and Detection}}, \href{https://doi.org/10.1393/ncr/i2016-10120-8}{\emph{Riv. Nuovo Cim.} {\bfseries 39} (2016) 1} [\href{https://arxiv.org/abs/1508.00785}{{\ttfamily 1508.00785}}].

\bibitem{Dreiner:2003wh}
H.~K. Dreiner, C.~Hanhart, U.~Langenfeld and D.~R. Phillips, \emph{{Supernovae and light neutralinos: SN1987A bounds on supersymmetry revisited}}, \href{https://doi.org/10.1103/PhysRevD.68.055004}{\emph{Phys. Rev. D} {\bfseries 68} (2003) 055004} [\href{https://arxiv.org/abs/hep-ph/0304289}{{\ttfamily hep-ph/0304289}}].

\bibitem{Dreiner:2013mua}
H.~K. Dreiner, J.-F. Fortin, C.~Hanhart and L.~Ubaldi, \emph{{Supernova constraints on MeV dark sectors from $e^+e^-$ annihilations}}, \href{https://doi.org/10.1103/PhysRevD.89.105015}{\emph{Phys. Rev. D} {\bfseries 89} (2014) 105015} [\href{https://arxiv.org/abs/1310.3826}{{\ttfamily 1310.3826}}].

\bibitem{Caputo:2021rux}
A.~Caputo, G.~Raffelt and E.~Vitagliano, \emph{{Muonic boson limits: Supernova redux}}, \href{https://doi.org/10.1103/PhysRevD.105.035022}{\emph{Phys. Rev. D} {\bfseries 105} (2022) 035022} [\href{https://arxiv.org/abs/2109.03244}{{\ttfamily 2109.03244}}].

\bibitem{Carenza:2023old}
P.~Carenza, G.~Lucente, L.~Mastrototaro, A.~Mirizzi and P.~D. Serpico, \emph{{Comprehensive constraints on heavy sterile neutrinos from core-collapse supernovae}},  \href{https://arxiv.org/abs/2311.00033}{{\ttfamily 2311.00033}}.

\bibitem{Pejcha:2015pca}
O.~Pejcha and J.~L. Prieto, \emph{{On The Intrinsic Diversity of Type II-Plateau Supernovae}}, \href{https://doi.org/10.1088/0004-637X/806/2/225}{\emph{Astrophys. J.} {\bfseries 806} (2015) 225} [\href{https://arxiv.org/abs/1501.06573}{{\ttfamily 1501.06573}}].

\bibitem{Muller:2017bdf}
T.~M\"uller, J.~L. Prieto, O.~Pejcha and A.~Clocchiatti, \emph{{The Nickel Mass Distribution of Normal Type II Supernovae}}, \href{https://doi.org/10.3847/1538-4357/aa72f1}{\emph{Astrophys. J.} {\bfseries 841} (2017) 127} [\href{https://arxiv.org/abs/1702.00416}{{\ttfamily 1702.00416}}].

\bibitem{Goldberg:2019ktf}
J.~A. Goldberg, L.~Bildsten and B.~Paxton, \emph{{Inferring Explosion Properties from Type II-Plateau Supernova Light Curves}}, \href{https://doi.org/10.3847/1538-4357/ab22b6}{\emph{Astrophys. J.} {\bfseries 879} (2019) 3} [\href{https://arxiv.org/abs/1903.09114}{{\ttfamily 1903.09114}}].

\bibitem{Murphy:2019eyu}
J.~W. Murphy, Q.~Mabanta and J.~C. Dolence, \emph{{A Comparison of Explosion Energies for Simulated and Observed Core-Collapse Supernovae}}, \href{https://doi.org/10.1093/mnras/stz2123}{\emph{Mon. Not. Roy. Astron. Soc.} {\bfseries 489} (2019) 641} [\href{https://arxiv.org/abs/1904.09444}{{\ttfamily 1904.09444}}].

\bibitem{SNprofile}
``Garching core-collapse supernova research archive.'' \url{https://wwwmpa.mpa-garching.mpg.de/ccsnarchive/}.

\bibitem{Bollig:2020xdr}
R.~Bollig, W.~DeRocco, P.~W. Graham and H.-T. Janka, \emph{{Muons in Supernovae: Implications for the Axion-Muon Coupling}}, \href{https://doi.org/10.1103/PhysRevLett.125.051104}{\emph{Phys. Rev. Lett.} {\bfseries 125} (2020) 051104} [\href{https://arxiv.org/abs/2005.07141}{{\ttfamily 2005.07141}}].

\bibitem{Chang:2016ntp}
J.~H. Chang, R.~Essig and S.~D. McDermott, \emph{{Revisiting Supernova 1987A Constraints on Dark Photons}}, \href{https://doi.org/10.1007/JHEP01(2017)107}{\emph{JHEP} {\bfseries 01} (2017) 107} [\href{https://arxiv.org/abs/1611.03864}{{\ttfamily 1611.03864}}].

\bibitem{Weldon:1983jn}
H.~A. Weldon, \emph{{Simple Rules for Discontinuities in Finite Temperature Field Theory}}, \href{https://doi.org/10.1103/PhysRevD.28.2007}{\emph{Phys. Rev. D} {\bfseries 28} (1983) 2007}.

\bibitem{Lucente:2022vuo}
G.~Lucente, L.~Mastrototaro, P.~Carenza, L.~Di~Luzio, M.~Giannotti and A.~Mirizzi, \emph{{Axion signatures from supernova explosions through the nucleon electric-dipole portal}}, \href{https://doi.org/10.1103/PhysRevD.105.123020}{\emph{Phys. Rev. D} {\bfseries 105} (2022) 123020} [\href{https://arxiv.org/abs/2203.15812}{{\ttfamily 2203.15812}}].

\bibitem{Diamond:2023scc}
M.~Diamond, D.~F.~G. Fiorillo, G.~Marques-Tavares and E.~Vitagliano, \emph{{Axion-sourced fireballs from supernovae}}, \href{https://doi.org/10.1103/PhysRevD.107.103029}{\emph{Phys. Rev. D} {\bfseries 107} (2023) 103029} [\href{https://arxiv.org/abs/2303.11395}{{\ttfamily 2303.11395}}].

\bibitem{Fischer:2016cyd}
T.~Fischer, S.~Chakraborty, M.~Giannotti, A.~Mirizzi, A.~Payez and A.~Ringwald, \emph{{Probing axions with the neutrino signal from the next galactic supernova}}, \href{https://doi.org/10.1103/PhysRevD.94.085012}{\emph{Phys. Rev. D} {\bfseries 94} (2016) 085012} [\href{https://arxiv.org/abs/1605.08780}{{\ttfamily 1605.08780}}].

\bibitem{Vogel:2013raa}
H.~Vogel and J.~Redondo, \emph{{Dark Radiation constraints on minicharged particles in models with a hidden photon}}, \href{https://doi.org/10.1088/1475-7516/2014/02/029}{\emph{JCAP} {\bfseries 02} (2014) 029} [\href{https://arxiv.org/abs/1311.2600}{{\ttfamily 1311.2600}}].

\bibitem{Sutherland:1975dr}
P.~Sutherland, J.~N. Ng, E.~Flowers, M.~Ruderman and C.~Inman, \emph{{Astrophysical Limitations on Possible Tensor Contributions to Weak Neutral Current Interactions}}, \href{https://doi.org/10.1103/PhysRevD.13.2700}{\emph{Phys. Rev. D} {\bfseries 13} (1976) 2700}.

\bibitem{Braaten:1993jw}
E.~Braaten and D.~Segel, \emph{{Neutrino energy loss from the plasma process at all temperatures and densities}}, \href{https://doi.org/10.1103/PhysRevD.48.1478}{\emph{Phys. Rev. D} {\bfseries 48} (1993) 1478} [\href{https://arxiv.org/abs/hep-ph/9302213}{{\ttfamily hep-ph/9302213}}].

\bibitem{Beck:2001dz}
D.~H. Beck and B.~R. Holstein, \emph{{Nucleon structure and parity violating electron scattering}}, \href{https://doi.org/10.1142/S0218301301000381}{\emph{Int. J. Mod. Phys. E} {\bfseries 10} (2001) 1} [\href{https://arxiv.org/abs/hep-ph/0102053}{{\ttfamily hep-ph/0102053}}].

\end{thebibliography}\endgroup

\end{document}